\documentclass{amsart}
\usepackage{amssymb,amscd}
\newcommand{\nc}{\newcommand}
\nc{\gl}{\mathfrak{gl}}
\nc{\WN}{\mathbf{N}}
\nc{\dBGG}{\dot{\BGG}}
\nc{\schi}{\bar{\chi}}
\nc{\hW}{\bh_{\W}}
\nc{\dualhW}{\bh^*_{\W}}
\nc{\bt}{\widehat{\t}}
\nc{\nat}{{\natural}}
\nc{\Prnongen}{{\textsl{Pr}}_{\kappa,\mathrm{nondeg}}}
\nc{\bgf}{\widehat{\g}^f}
\nc{\bge}{\widehat{\g}^e}
\nc{\rootsW}{\widehat{\roots}^{\t}}
\nc{\sgf}{\sg^f}
\nc{\sge}{\sg^e}
\nc{\QW}{\widehat{Q}^{\t}}
\nc{\Center}{{\mathcal{Z}}}
\nc{\Wcat}{{\mathbb{O}}_k(\sg,f)}
\nc{\fmap}{\rightsquigarrow }
\nc{\fdomain}{{\mathcal{C}}}
\nc{\Hecke}{{\mathcal{H}}}
\nc{\Neu}{U(\n_{\ell})_{\chi}}
\nc{\D}{{\mathbf{D}}}
\nc{\mg}{\g^{\flat}}
\nc{\cat}{{\mathcal{C}}}
\nc{\broots}{\widehat{\roots}}
\nc{\tri}{\triangle}
\nc{\Lg}{L^{\g}}
\nc{\delW}{\delta^{\W}}

\newcommand{\HW}{{{S}}}

\newcommand{\teigi}{\underset{\mathrm{def}}{=}}

\newcommand{\why}{{\mathbf{L}}}

\newcommand{\M}{{\mathbf{M}}}

\newcommand{\U}{{\mathcal{U}}}
\newcommand{\DW}{{\bf D}^{\mathcal W}}

\newcommand{\Dg}{{\bf D}}

\newcommand{\W}{{\mathcal{W}}}

\renewcommand{\t}{{\mathfrak{t}}}

\newcommand{\1}{{\mathbf{1}}}

\newcommand{\obj}{Obj}

\newcommand{\sroots}{{\Delta}}
\newcommand{\roots}{\Delta}

\newcommand{\F}{{\mathcal{F}}}

\newcommand{\sa}{{\mathfrak{a}}}
\renewcommand{\sl}{{\mathfrak{sl}}}

\newcommand{\Cl}{{\mathcal{C}l}}

\newcommand{\semiinf}{\frac{\infty}{2}}

\newcommand{\C}{{\mathbb C}}
\newcommand{\Z}{{\mathbb Z}}

\newcommand{\inv}{^{-1}}
\newcommand{\dual}[1]{{#1}^*}
\newcommand{\lam}{\lambda}
\newcommand{\Lam}{\Lambda}
\renewcommand{\P}{{\mathcal P}}
\renewcommand{\*}{{\otimes}}
\newcommand{\+}{\mathop{\oplus}}
\newcommand{\h}{ {\mathfrak h}}
\newcommand{\g}{ {\mathfrak g}}

\newcommand{\che}{^{\vee}}

\newcommand{\bra}{{\langle}}
\newcommand{\ket}{{\rangle}}

\DeclareMathOperator{\sdim}{sdim}
\DeclareMathOperator{\str}{str}

\DeclareMathOperator{\neu}{ne}

\DeclareMathOperator{\End}{End}

\DeclareMathOperator{\Hom}{Hom}

\DeclareMathOperator{\id}{id}
\DeclareMathOperator{\ad}{ad}
\DeclareMathOperator{\haru}{span}

\DeclareMathOperator{\ch}{ch}

\DeclareMathOperator{\gr}{gr}

\newcommand{\sg}{  {\mathfrak g}}
\newcommand{\sh}{{ \h}}
\newcommand{\sn}{{\mathfrak{n}}}

\newcommand{\srho}{ \bar{\rho}}

\newcommand{\sproots}{{\Delta}_+}

\newcommand{\bg}{\widehat{\g}}
\newcommand{\bh}{ \widehat{\h}}
\newcommand{\bn}{\widehat{\n}}

\newcommand{\bQ}{ \widehat{Q}}

\newcommand{\ud}[2]{{\genfrac{}{}{0pt}{}{#1}{#2}}}

\newcommand{\n}{{{\mathfrak{n}}}}

\newcommand{\BGG}{{\mathcal O}}

\newcommand{\Obj}{Obj}

\title[Representation theory of superconformal algebras]{
Representation theory of superconformal algebras
and the Kac-Roan-Wakimoto conjecture
}
\author{Tomoyuki Arakawa}
\address{Graduate school of Mathematics,  Nagoya
University,
Chikusa-ku, Nagoya, 464-8602, JAPAN}
\email{tarakawa@math.nagoya-u.ac.jp}
\begin{document}
\theoremstyle{plain}
\newtheorem{Th}{Theorem}[subsection]
\newtheorem*{MainTh}{Main Theorem}
\newtheorem{Pro}[Th]{Proposition}
\newtheorem{ProDef}[Th]{Proposition and Definition}
\newtheorem{Lem}[Th]{Lemma}
\newtheorem{Co}[Th]{Corollary}

\newtheorem{Facts}[Th]{Facts}

\theoremstyle{definition}

\newtheorem{dfandpr}[Th]{Definition and Proposition}
\theoremstyle{remark}
\newtheorem{Def}[Th]{Definition}
\newtheorem{Rem}[Th]{Remark}
\newtheorem{Conj}{Conjecture}
\newtheorem{Claim}{Claim}
\newtheorem{Notation}{Notation}
\newtheorem{Ex}[Th]{Example}

\newcommand{\st}{{\mathrm{st}}}
\maketitle
\begin{abstract}
We study
the representation theory  of the
superconformal algebra $\W_k(\sg,f_{\theta})$
associated with a minimal gradation of $\g$.
Here, $\sg$ is a simple finite-dimensional Lie superalgebra
with a non-degenerate, even supersymmetric invariant bilinear form.
Thus,
$\W_k(\sg,f_{\theta})$ can be 
one of the well-known superconformal algebras
including
the Virasoro algebra,
the Bershadsky-Polyakov algebra,
the Neveu-Schwarz algebra,
the Bershadsky-Knizhnik algebras,
the $N=2$ superconformal algebra,
the $N=4$ superconformal algebra,
the $N=3$ superconformal algebra
and the big $N=4$ superconformal algebra.
We prove
the conjecture of
V.\ G.\ Kac,
S.-S.\ Roan and M. Wakimoto for $\W_k(\sg,f_{\theta})$.
In fact,
we  show that
any irreducible highest weight character 
of $\W_k(\sg,f_{\theta})$
at any level $k\in \C$
is determined by the
corresponding
irreducible highest weight character 
of the Kac-Moody affinization of  $\sg$.
 \end{abstract}
\setcounter{tocdepth}{1}
\section{Introduction}
Suppose that the
following are is given:
(i) a simple finite-dimensional Lie superalgebra 
$\sg$ with non-degenerate, even supersymmetric invariant bilinear form,
(ii) a nilpotent element $f$ in the even part of $\sg$
and (iii) a level $k\in \C$ of 
 the Kac-Moody affinization $\bg$ of $\sg$.
Then,
the corresponding
 $\W$-algebra $\W_k(\sg,f)$
can be constructed using the method of
quantum BRST reduction.
This method 
was first introduced by B. L. Fe\u{\i}gin and E. V. Frenkel \cite{FF_W, FF_W2}
in the case that $\sg$ is a Lie algebra
and $f$ is its principal nilpotent element,
and it was
recently 
extended
 to the general case 
by V. G. Kac, S.-S. Roan and M. Wakimoto \cite{KRW}.

In this paper we study the representation theory 
of
those $\W$-algebras for which the nilpotent element
$f$
is
equal to
the root vector 
$f_{\theta}$
 corresponding to the lowest root $-\theta$ of $\sg$.
V. G. Kac, S.-S. Roan and M. Wakimoto \cite{KRW}
showed that these $\W$-algebras $\W_k(\sg,f_{\theta})$
are quite different from  the standard $\W$-algebras
\cite{Za,FZa,  Luk,FL, LF, FF_W, FF_W2}
associated with principal nilpotent elements
and noteworthy because
they include almost all the superconformal algebras
so far constructed to this time by physicists
such as 
the Virasoro algebra,
the Bershadsky-Polyakov algebra \cite{B},
the Neveu-Schwarz algebra,
the Bershadsky-Knizhnik algebras \cite{BeK},
the $N=2$ superconformal algebra,
the $N=4$ superconformal algebra,
the $N=3$ superconformal algebra and
the big $N=4$ superconformal algebra.

\smallskip

Let
$\BGG_k$
be the 
Bernstein-Gel'fand-Gel'fand category of
$\bg$ at level $k\in \C$.
Let $M(\lam)$ be the Verma module of $\bg$
with highest weight $\lam$
and
$L(\lam)$ be  the unique irreducible
quotient of $M(\lam)$.
The method of
quantum BRST reduction
gives 
 a family of functors $V\fmap H^i(V)$,
depending on $i\in \Z$,
from $\BGG_k$ to the category of
$\W_k(\sg,f)$-modules.
Here,
$H^{\bullet}(V)$
is the BRST cohomology of the corresponding 
quantum reduction.
In the case that $\sg$ is a Lie algebra
and $f$ is its principal nilpotent element,
E. V. Frenkel, V. G. Kac and M. Wakimoto \cite{FKW}
used this functor in their construction of the
``minimal'' series presentations of $\W_k(\sg,f)$,
and 
they conjectured that 
$H^{\bullet}(L(\lam))$
is irreducible (or zero)
for an admissible
weight $\lam$.
This conjecture 
was extended  by V.\ G.\ Kac, S.-S.\ Roan and M.\ Wakimoto
 \cite{KRW}
to the general case,
in which
they conjectured that,
for an admissible
weight $\lam$,
the irreduciblity of 
$H^{\bullet}(L(\lam))$
holds
for a general pair $(\g,f)$
consisting
of a  Lie superalgebra $\sg$
and a nilpotent element $f$ 
(see Conjecture 3.1B of Ref.\ \cite{KRW}).

As a  continuation of the present author's previous work \cite{A1,A2},
in which
the  conjecture of E. V. Frenkel, V. G. Kac and M. Wakimoto 
 was proved
  (completely for the ``$-$'' case
and partially for the ``$+$'' case),
we prove the conjecture
of V.\ G.\ Kac, S.-S.\ Roan and M.\ Wakimoto
for $\W_k(\sg,f_{\theta})$;
Actually we prove even stronger results,
showing
that
the
representation theory of 
$\W_k(\sg,f_{\theta})$
is 
controlled 
by   $\bg$
in the following sense:

\smallskip

\begin{MainTh}
For arbitrary level $k\in \C$,
we have the following:

\begin{enumerate}
 \item \rm{(Theorem \ref{Th:vanishin})}
 We have
$H^i(V)=\{0\}$ with $i\ne 0$
for any $V\in \Obj\BGG_k$.
\item \rm{(Theorem \ref{Th;image-of-simples})}
Let $L(\lam)\in \obj\BGG_k$
be the irreducible $\bg$-module with highest weight
$\lam$.
Then,
 $\bra \lam,\alpha_0\che\ket
\in\{0,1,2,\dots\}$ implies
$H^0(L(\lam))=\{0\}$.
Otherwise,
$H^0(L(\lam))$ is isomorphic to the irreducible 
$\W_k(\sg,f_{\theta})$-module
with the corresponding highest weight.
\end{enumerate}
\end{MainTh}

 Main Theorem  (1)  implies, in particular, that
the correspondence  $V\fmap H^0(V)$
defines an exact functor from $\BGG_k$
to the category of $\W_k(\sg,f_{\theta})$-modules,
defining a map between characters. 
On the other hand,
every 
irreducible highest weight representation
of $\W_k(\sg,f_{\theta})$
is isomorphic to $H^0(L(\lam))$
for some $\lam$ (see Section \ref{section:Irr-HW}).
It is also known that $H^0(M(\lam))$
is a Verma module over $\W_k(\sg,f_{\theta})$ \cite{KW2003}.
Hence, 
from the above results,
 it follows
  that
the character of
any irreducible highest weight representation
of $\W_k(\sg,f_{\theta})$
is determined by the
character of the 
corresponding 
irreducible $\bg$-module
$L(\lam)$.
%
%
%
We remark that 
Main Theorem (2)
is consistent  with the  computation 
of V. G. Kac, S.-S.\ Roan and M. Wakimoto \cite{KRW}
of the Euler-Poincar\'{e}
character of $H^{\bullet}(L(\lam))$.

\medskip

\medskip


\medskip

This paper is organized as follows.
In Section 2
we collect the necessary information regarding the affine Lie superalgebra
$\bg$.
In this setting,
a slight modification is need for the $A(1,1)$ case,
which is summarized in Appendix.
In Section 3
we recall the definition 
of  the  BRST complex
constructed by V. G. Kac, S.-S.\ Roan,
and M. Wakimoto \cite{KRW}.
As explained in Ref.\ \cite{KRW},
their main idea 
in generalizing the construction of B. L. Fe\u{\i}gin and E. V. Frenkel
\cite{FF_W,FF_W2}
was to  add  the ``neutral free superfermions''
whose definition is given at the beginning of that section.
Although the $\W$-algebra
$\W_k(\sg,f)$
can be defined for an arbitrary 
nilpotent element $f$,
the assumption $f=f_{\theta}$
simplifies the theory in many ways.
This is also the case that 
the above-mentioned well-known superconformal algebras appear,
as explained in Ref.\ \cite{KRW}.
In Section 4
we derive some basic but important facts concerning the BRST 
cohomology under
the assumption $f=f_{\theta}$.
In Section 5  we recall the definition of the $\W$-algebra
$\W_k(\sg,f)$
and collect necessary information about its structure.
In Section 6
we present the parameterization
of irreducible highest weight 
representations of $\W_k(\sg,f_{\theta})$
and
 state our main results (Theorems 
 \ref{Th:vanishin}
and \ref{Th;image-of-simples}).
The most important part of the proof
is the  computation
of
the BRST cohomology 
$H^{\bullet}(M(\lam)^*)$
associated with the dual $M(\lam)^*$
of the Verma module $M(\lam)$.
This is carried out in Section 7
by introducing
 a particular spectral sequence.
The argument used here is a modified version of
that given in Ref.\ \cite{A1},
where we proved the vanishing of the cohomology
associated to the original quantum reduction 
formulated by B. L. Fe\u{\i}gin and E. V. Frenkel
\cite{FF_W, FF_W2, FKW}.

\smallskip
The method used in this paper can  also be applied to
general $\W$-algebras,
with some modifications.
Our results
for that case
 will appear in  forthcoming papers.

\bigskip

\noindent{\em Acknowledgments.}
This work was started
during my visit to M.I.T.,  from February to March 2004.
I would like to thank the people of M.I.T.
for their hospitality,
and in particular
Professor Victor G. Kac .
Also,
I would like to thank the anonymous referees for their 
very helpful suggestions.

\section{Preliminaries}
\label{section:Preliminaries}
\subsection{}
Let
$\sg$
be a complex, simple finite-dimensional
Lie superalgebra with a  non-degenerate,
even supersymmetric invariant
bilinear form $(.|.)$.
Let
 $(e,x,f)$ be 
a $\mathfrak{sl}_2$-triple
in the even part of $\sg$
normalized as 
 \begin{align}\label{def-of-sl2}
 [e,f]=x,\
[x,e]=e,
[x,f]=-f.
\end{align}
Further, let
\begin{align}\label{eq:fun-gradation}
 \text{$\sg=\bigoplus\limits_{j\in \frac{1}{2}\Z}\sg_j$},
\quad \sg_j=\{u\in \g \mid [x,u]=j a\}
\end{align}
be the eigenspace decomposition of $\sg$ with respect to $\ad x$.

\subsection{}
Define
$\sg^f$ as the centralizer of $f$ in $\sg$,
so that
$\sgf=\{u\in  \sg\mid [f,u]=0\}$.
Then,
we have
$\sg^f=\sum_{j\leq 0}\sg_j^f$,
where $\sg^f_j=\sg^f \cap \sg_j$.
Similarly, we define
$\sg^e\teigi
\{u\in \sg\mid [e,u]=0\}=\sum_{j\geq 0}\sg_j^e$.
\subsection{}
Define the following:
\begin{align*}
&\sg_{\geq 1}\teigi\bigoplus_{j\geq 1}\sg_j,\quad
 \sg_{>0}\teigi\bigoplus_{j>0}\sg_{j}.
\end{align*}
Then, $\sg_{>0}=\sg_{\geq 1}\+ \sg_{\frac{1}{2}}$,
and
$\sg_{\geq 1}$ and $\sg_{>0}$ are both nilpotent subalgebras of $\sg$.
Similarly define 
$\sg_{\geq 0}$,
$\sg_{\leq 0}$,
$\sg_{<0}$
and
$\sg_{\leq -1}$.
\subsection{}
Define 
the character $\bar{\chi}$ of 
$\sg_{\geq 1}$
by
\begin{align}
 \bar{\chi}(u)\teigi ( f|u),\quad \text{where $u\in \sg_{\geq 1}$.}
\end{align}
Then,
$\schi$
defines a
skew-supersymmetric even bilinear form 
$\bra . | . \ket_{\neu}$ 
on 
${\sg}_{\frac{1}{2}}$
through the formula 
\begin{align}
 \text{$\bra u|v\ket_{\neu}
=\bar{\chi}([u,v])$.}
\end{align}
Note that
$\bra . | . \ket_{\neu}$ 
is non-degenerate,
as guaranteed
by the $\mathfrak{sl}_2$-representation theory.
Also, we have the property
\begin{align}
 \bra u|[a,v]\ket_{\neu}=\bra [u,a]|v\ket_{\neu}
\quad \text{for $a\in \sg_0^f, u,v\in \sg_{\frac{1}{2}}$.}
\end{align}
\subsection{}
Let
$\sh\subset \sg_0$
be a Cartan subalgebra of $\sg$
containing $x$,
and
let
$\roots\subset \dual{\sh}$
be the set of roots.
It is known that
the root space $\sg^{\alpha}$,
with $\alpha\in \sroots$,
is one-dimensional 
except for $A(1,1)$ (see \cite{KacLNM}).
For this reason,
in the case of $A(1,1)$,
a slight modification is needed in the following
argument,
which is summarized in Appendix.
Now, let 
$\roots_j=\{\alpha\in \roots\mid \bra \alpha,x\ket=j\}$
with $j\in \frac{1}{2}\Z$.
Then,
$\sroots=\sqcup_{j\in \frac{1}{2}\Z}
\sroots_j$,
and $\roots_0$
is the set of roots of $\sg_0$.
Let $\sroots_{0+}$
be a set of positive roots of $\sroots_0$.
Then,
 $\sproots=\sroots_{0+}\sqcup
\sroots_{>0}$
is a set of positive roots of $\sg$,
where
$\sroots_{>0}=\sqcup_{j>0}\sroots_j$.
This gives the triangular decompositions
\begin{align}
 \sg=\sn_-\+ \sh\+ \sn_+,\quad
\sg_0=\sn_{0,-}\+\sh\+ \sn_{0,+}.
\end{align}
Here,
$\sn_{+}\teigi\sum_{\alpha\in \sproots}\sg_{\alpha}$,
$\sn_{0,+}\teigi\sum_{\alpha\in \sroots_{0,+}}\sg_{\alpha}$
and analogously for $\sn_-$ and $\sn_{0,-}$.
\subsection{}
Let $u\mapsto u^t$ be an anti-automorphism of $\sg$
such that $e^t=f$,
$f^t=e$,
$x^t=x$,
$\sg_{\alpha}^t=\sg_{-\alpha}$
for
$\alpha\in \sroots$
and
$(u^t|v^t)=(v|u)$ for $u,v\in\sg$.
We fix the root vectors
$u_{\alpha}\in \sg_{\alpha}$,
where
$\alpha\in \sroots$,
such that
$(u_{\alpha},u_{-\alpha})=1$
and $u_{\alpha}^t=u_{-\alpha}$
with $\alpha\in \sproots$.
\subsection{}
Let $p(\alpha)$ be the parity
of  $\alpha\in \sroots$ and
$p(v)$ be the
parity of $v\in \sg$.
\subsection{}
Let
$\bg$
be the {\em Kac--Moody affinization} of $\sg$.
This is the Lie superalgebra
given by
\begin{align}\label{eq-def-of-affine-Lie-algebra}
 \bg=\sg\* \C[t,t\inv]\+ \C K\+\C \D
\end{align}
with the commutation relations
\begin{align}
& [u(m),v(n)]=[u,v](m+n)+m\delta_{m+n,0}(u|v)K,\\
&[\D,u(m)]=m u(m),\quad [K,\bg]=0
\end{align}
for $u,v\in \sg$, $m,n\in \Z$.
Here,
$u(m)\teigi u\* t^m$
 for $u\in \sg$ and
$m\in \Z$.

The invariant bilinear form
$(\cdot|\cdot)$
is extended from $\sg$
to $\bg$
by stipulating
$(u(m)|v(n))=(u|v)\delta_{m+n,0}$ 
with $u,v\in \sg$, $m,n\in \Z$,
$(\sg\* \C[t,t\inv],\C K\+ \C \D)=0$,
$(K,K)=(\D,\D)=0$
and $(K,\D)=(\D,K)=1$.

\subsection{}
Define the  subalgebras
\begin{align}
 L\sg_{\geq 1}\teigi\sg_{\geq 1}\* \C[t,t\inv],\quad
L\sg_{>0}\teigi\sg_{>0}\* \C[t,t\inv]\subset \bg.
\end{align}
Similarly, define 
$L\sg_{\geq 0}$,
$L\sg_{\leq 0}$,
$L\sg_{<0}$ and
$L\sg_{\leq -1}$.

\subsection{}
Fix the triangular decomposition
$\bg=\bn_-\+ \bh\+\bn_+$
in the standard way.
Then, we have
\begin{align*}
&\text{$\bh=\sh\+ \C K\+ \C \D$},\\
&\bn_-=\sn_-\*\C[t\inv]\+ \sh\* \C[t\inv]t\inv\+\sn_+\*\C[t\inv]t\inv,\\
&\bn_+=\sn_-\* \C[t]t\+ \sh\*\C[t]t\+ \sn_+\* \C[t].
\end{align*}
Let
$\dual{\bh}=\dual{\sh}\+ \C \Lam_0 \+ \C \delta$
be the dual of $\bh$,
where,
$\Lam_0$ and $\delta$ are dual elements of $K$ and $\D$, respectively.
Next, let
$\broots$
be the set of roots of $\bg$,
$\broots_+$ the set of positive roots,
and $\broots_-$ the set of negative roots.
Then, we have
$\broots_-=-\broots_+$.
Further,
let
$\bQ$ be the root lattice and 
$\bQ_+=\sum_{\alpha\in \broots_+}\Z_{\geq 0}
\alpha\subset \bQ$.
We define a partial 
ordering $\mu\leq  \lam$  on $\dual{\bh}$
by
$\lam-\mu\in \bQ_+$.

\subsection{}
For  an $\bh$-module $V$,
let $V^{\lam}$ be the weight space of weight $\lam$,
that is,
$V^{\lam}\teigi\{v\in V\mid
hv=\lam(h)v\text{ for all }h\in \bh\}$.
If all the weight spaces $V^{\lam}$ are finite-dimensional,
we define
the graded dual $V^*$ of $V$
by
\begin{align}\label{eq:def-of-dual}
 \dual{V}\teigi\bigoplus_{\lam\in \dual{\h}}\Hom_{\C}(V^{\lam},\C)
\subset \Hom_{\C}(V,\C).
\end{align}

\subsection{}\label{subsection:category-O}
Throughout this paper, 
$k$ represents a complex number.
Let $\dual{\bh}_{k}$
denote the set of weights of $\bg$ of level $k$:
\begin{align}
 \dual{\bh}_k\teigi\{ \lam\in \dual{\bh}\mid
\bra \lam,K\ket=k\}.
\end{align}
Also, let $\BGG_k$
be the full subcategory of the category of left $\bg$-modules consisting
of objects $V$
such that the following hold:
\begin{enumerate}
 \item $V=\bigoplus_{\lam\in \dual{\bh}_k}V^{\lam}$
and $\dim_{\C}V^{\lam}<\infty$ for all $\lam\in \dual{\bh}_k$;
\item there exists a finite set
$\{\mu_1,\dots,\mu_r\}\subset \dual{\bh}_k$ such that
$\lam\in \bigcup_i (\mu_i-\bQ_+)$ for any weight  $\lam$
with $V^{\lam}\ne \{0\}$.
\end{enumerate}
Then,
$\BGG_k$ is an abelian category.
Let
$M(\lam)\in \Obj\BGG_k$,
with
$\lam\in \dual{\bh}_k$,
be the Verma module with highest weight $\lam$.
That is,
$M(\lam)=U(\bg)\*_{U(\bh\+ \bn_+)}\C_{\lam}$,
where $\C_{\lam}$
is the one-dimensional $\bh\+\bn_+$-module on which
$\bn_+$ acts trivially and $h\in \bh$
acts as $\bra \lam,h\ket \id$.
Let $v_{\lam}$ be the highest weight vector of $M(\lam)$.
Next, let
 $L(\lam)\in \Obj\BGG_k$ be the  unique simple quotient of $M(\lam)$. 

\subsection{}
The correspondence $V\fmap \dual{V}$ defines the duality functor
in $\BGG_k$.
Here,
$\bg$ acts on $\dual{V}$
as $(a f)(v)=f(a^t v)$,
where $a\mapsto a^t$
is the antiautomorphism of $\bg$
defined by
$u(m)^t= (u^t)(-m)$
(with $u\in \sg,m\in \Z$),
$K^t=K$
and
$\D^t=\D$.
We have $L(\lam)^*=L(\lam)$
for $ \lam\in \dual{\bh}_k$.
\subsection{}
Let
$\BGG_k^{\tri}$
be the full subcategory of $\BGG_k$
consisting of objects $V$ that admit a Verma flag,
that is,
a finite filtration 
$V=V_0\supset V_1\supset \dots \supset V_r=\{0\}$
such that each successive subquotient $
V_i/V_{i+1}$
is isomorphic to
some Verma module $M(\lam_i)$ with $\lam_i\in \dual{\bh}$.
The category $\BGG_k^{\tri}$
is stable under the operation of taking a direct summand.
Dually,
let $\BGG_k^{\bigtriangledown}
$  be 
the full subcategory of $\BGG_k$
consisting of objects $V$ such that
$\dual{V}\in \Obj\BGG^{\tri}_k$.

\subsection{}
For $\lam\in \dual{\bh}_k$,
let
$\BGG_k^{\leq \lam}$
be the full subcategory of $\BGG_k$
consisting of objects $V$ 
such that
$V=\bigoplus\limits_{\mu\leq \lam}V^{\mu}$.
Then,
$\BGG_k^{\leq \lam}$
is an abelian category and stable under the operation of taking 
(graded) dual.
Also,
every simple object $L(\mu)\in \Obj\BGG_k^{\leq \lam}$,
with $\mu\leq \lam$,
admits a projective
cover $P_{\leq \lam}(\mu)$
in $\BGG_{k}^{\leq \lam}$,
and hence,
every finitely generated object in $\BGG_k^{\leq \lam}$
is an image of a projective object
of the form
$\bigoplus_{i=1}^{r}
P_{\leq \lam}(\mu_i)$ for some $\mu_i\in \dual{\bh}$.
Indeed,
as in the Lie algebra case (see, e.g., Ref.\ \cite{MPBOOK}),
$P_{\leq \lam}(\mu)$
can be defined as
 an indecomposable direct summand
of
\begin{align*}
 U(\bg)\*_{U(\bh\+ \bg_+)}\tau_{\leq \lam}\left(
U(\bh\+ \bn_+)\*_{U(\bh)}\C_{\mu}\right)
\end{align*}
that has $L(\mu)$ as a quotient.
Here,
$\tau_{\leq \lam}(V)=V/\bigoplus\limits_\ud{\nu\in \dual{\bh}
}{\nu\not\leq \lam}V^{\nu}$,
and $\C_{\mu}$ is a one-dimensional $\bh$-module
on which $h\in \bh$ acts as $\mu(h)\id$.
Note that
$P_{\leq \lam}(\mu)\in \Obj \BGG_k^{\tri}$.
Moreover,
the Bernstein-Gel'fand-Gel'fand reciprocity
holds:
\begin{align*}
 [P_{\leq \lam}(\mu):M(\mu')]
=[M(\mu')
:L(\mu)]
\quad \text{(with $\mu,\mu'\leq \lam$).}
\end{align*}
Here,
$[P_{\leq \lam}(\mu):M(\mu')]$
is the multiplicity of $M(\mu')$
in the Verma flag of $P_{\leq \lam}(\mu)$,
and $[M(\mu'):L(\mu)]$
is the multiplicity of $L(\mu)$
in the local composition factor (see Ref.\ \cite{KacBook}) of $M(\mu')$.
Dually,
$I_{\leq \lam}(\mu)=P_{\leq \lam}(\mu)^*$
is
the injective envelope of $L(\mu)$
in $\BGG_k^{\leq \lam}$.
In particular,
 $V\in \Obj\BGG_k^{\leq \lam}$
is a submodule of an injective object of the form
$\bigoplus_{i=1}^r
I_{\leq \lam}(\mu_i)$
for some $\mu_i\in \dual{\bh}$
if its dual $V^*$ is
finitely generated.

\section{The Kac-Roan-Wakimoto construction I: the BRST complex}
\subsection{}
Define a character $\chi$
of $L\sg_{\geq 1}$
by
\begin{align}
 \chi (u(m))\teigi (f(1)|u(m))=\bar{\chi}(u)\delta_{m,-1}\quad
\text{for $u\in \sg_{\geq 1}, m\in \Z$.}
\end{align}
Let $\ker \chi\subset U(L\sg_{\geq 1})$
be the kernel of the algebra homomorphism $\chi: U(L\sg_{\geq 1})
\rightarrow \C$.
Define
$I_{\chi}\teigi U(L\sg_{>0}) \ker \chi$.
Then,
$I_{\chi}
$ is a two-sided ideal of $U(L\sg_{>0})$.
Next, define
\begin{align}\label{eq:def-of-N-chi-+}
 N(\chi)\teigi U(L\sg_{>0})/I_{\chi}.
\end{align}
Now, let
$\Phi_u(n)$,
with $u\in \sg_{>0}$ and
$n\in \Z$, denote the image of $u(n) \in  L\sg_{>0}$
in the algebra $N(\chi)$.
With a slight abuse of notation,
we write
$\Phi_{u_{\alpha}}(n)$ as
$ \Phi_{\alpha}(n)$
for $\alpha\in \sroots_{\frac{1}{2}}$ and $n\in \Z$.
Then,
the superalgebra
$N({\chi})$
is generated by
$\Phi_{\alpha}(n)$,
where $\alpha \in \sroots_{\frac{1}{2}}$ and $n\in \Z$,
with the relations
\begin{align}\label{eq:comm-rel-of-neu}
 [\Phi_{\alpha}(m), \Phi_{\beta}(n)]
=\bra u_{\alpha}|u_{\beta}\ket_{\neu}\delta_{m+n,-1}
\quad \text{for $\alpha,\beta\in \sroots_{\frac{1}{2}}, m,n\in \Z$.}
\end{align}
 The elements
$\{ \Phi_{\alpha}(n)\}$
are called
the {\em neutral free superfermions} (see Example 1.2 of Ref.\ \cite{KRW}).

Let $\{u^{\alpha}\}_{\alpha\in \sroots_{\frac{1}{2}}}$
be the basis of $\sg_{\frac{1}{2}}$
dual to $\{u_{\alpha}\}_{\alpha\in \sroots_{\frac{1}{2}}}$
with respect to $\bra ~|~\ket_{\neu}$,
that is,
$\bra u_{\alpha}|u^{\beta}\ket_{\neu}=\delta_{\alpha,\beta}$.
We set
$\Phi^{\alpha}(n)=\Phi_{u^{\alpha}}(n)$
for 
$\alpha\in \sroots_{\frac{1}{2}}$ and
$n\in \Z$,
and thus,
\begin{align}\label{eq:comm-rel-of-neu2}
 \text{$[\Phi_{\alpha}(m),\Phi^{\beta}(n)]
=\delta_{\alpha,\beta}\delta_{m+n,-1}$}
\quad \text{with $\alpha,\beta\in \roots_{\frac{1}{2}},\ m,n\in \Z$.}
\end{align}

\subsection{}
Let
$\F^{\neu}(\chi)$
be the irreducible representations of $N(\chi)$
generated by the vector $\1_{\chi}$
with the property
\begin{align}
 \Phi_{\alpha}(n) \1_{\chi}=0\quad 
\text{for $\alpha\in \sroots_{\frac{1}{2}}$ and $ n\geq 0$.}
\end{align}
Note that
$\F^{\neu}(\chi)$
is naturally a $L\sg_{>0}$-module through the 
algebra homomorphism $L\sg_{>0}\ni u(m)\mapsto \Phi_u(m)
\in N(\chi)$.

There is  a unique semisimple action of $\bh$ on
$\F^{\neu}(\chi)$
 such that the following hold:
\begin{align*}
& \text{$h\1_{\chi}=0$}\quad \text{for }h\in \sh,
\\ &\text{$\Phi_{\alpha}(n)\F^{\neu}(\chi)^{\lam}\subset 
\F^{\neu}(\chi)^{\lam+\alpha+n\delta}$
\quad \text{for }$\alpha\in \sroots_{\frac{1}{2}}$,
$n\leq -1$
\text{ and }
$\lam\in \dual{\bh}$.
}
\end{align*}
\begin{Lem}\label{Lem:strange-def-of-weights}
We have
$\Phi_{\alpha}(n)\F^{\neu}(\chi)^{\lam}
\subset
\sum\limits_{\beta\in \sroots_{\frac{1}{2}}\atop
 \schi([u_{\beta},u_{\alpha}])
\ne 0}
 \F^{\neu}(\chi)^{\lam-\beta+(n+1)\delta}$
 for $\alpha\in \sroots_{\frac{1}{2}}$, $n\geq 0$ and $\lam\in \dual{\bh}$.
\end{Lem}
\begin{proof}
By  definition we have
\begin{align}\label{eq:weird-action}
 \text{$\Phi^{\alpha}(n) \F^{\neu}(\chi)^{\lam}\subset
 \F^{\neu}(\chi)^{\lam-\alpha+(n+1)\delta}$
for 
$n\geq 0$
}
\end{align}
(see \eqref{eq:comm-rel-of-neu2}).
But $\Phi_{\alpha}(n)= \sum_{\beta\in \sroots_{\frac{1}{2}}}
\schi([u_{\beta},u_{\alpha}])\Phi^{\beta}(n)$ for $\alpha\in \sroots_{\frac{1}{2}}$.
\end{proof}
\subsection{}
Let
$\Cl(L\sg_{> 0})$ be the {\em Clifford superalgebra}
(or the {\em charged free superfermions})
associated with 
$L\sg_{>0}\+ \dual{(L\sg_{>0})}
$
and its natural bilinear from.
The superalgebra
$\Cl(L\sg_{>0})$
is generated by
$\psi_{\alpha}(n)$
and
$\psi^{\alpha}(n)$,
where $\alpha\in \sroots_{>0}$ and
$n\in \Z$,
with the relations
\begin{align*}
 &[\psi_{\alpha}(m),\psi^{\beta}(n)]=\delta_{\alpha,\beta}\delta_{m+n,0},\\
&[\psi_{\alpha}(m),\psi_{\beta}(m)]=[\psi^{\alpha}(m),\psi^{\beta}(n)]=0,
\end{align*}
where the parity of $\psi_{\alpha}(n)$
and $\psi^{\alpha}(n)$ is reverse to $u_{\alpha}$.
\subsection{}
Let $\F(L\sg_{>0})$ be the 
irreducible representation of
$\Cl(L\sg_{>0})$
generated by the vector $\1$
 that satisfies the relations
\begin{align*}
& \psi_{\alpha}(n)\1=0\quad 
\text{for $\alpha\in \sroots_{>0}$ and $n\geq 0$,}\\
\text{and }& \psi^{\alpha}(n)\1=0\quad 
\text{for $\alpha\in \sroots_{>0}$ and $ n> 0$.}
\end{align*}
The space
$\F(L\sg_{>0})$
is graded; that is,
$\F(L\sg_{>0})
=\bigoplus_{i\in \Z}\F^i(L\sg_{>0})$,
where  the degree is determined  from the assignments
$\deg \1=0$,
$\deg \psi_{\alpha}(n)=-1$
and $\deg \psi^{\alpha}(n)=1$,
with $\alpha\in \sroots_{>0}$ and
$n\in \Z$.

There is a natural 
semisimple action of $\bh$ on 
$\F(L\sg_{>0})$,
namely,
$\F(L\sg_{>0})=\bigoplus_{\lam\in \dual{\bh}}\F(L\sg_{>0})^{\lam}$.
This is defined by the relations
$h\1=0$
$(h\in \bh)$,
$\psi_{\alpha}(n)\F(L\sg_{>0})^{\lam}
\subset \F(L\sg_{>0})^{\lam+\alpha+n\delta}
$,
$\psi^{\alpha}(n)\F(L\sg_{>0})^{\lam}\subset
\F(L\sg_{>0})^{\lam-\alpha+n\delta}$,
where
$\alpha\in \sroots_{>0}$ and $n\in \Z$.
 \subsection{}
For $V \in \Obj \BGG_k$,
define
\begin{align}
 C(V)
\teigi V\* \F^{\neu}(\chi)\* \F(L\sg_{>0})
=\sum_{i\in \Z}C^i(V),
\end{align}
where 
$C^i(V)\teigi V\* \F^{\neu}(\chi)\* \F^i(L\sg_{>0})$.

Let
$\bh$ act on $C(V)$
by the tensor product action.
Then,
we have
$C(V)=\bigoplus_{\lam\in \dual{\bh}}C(V)^{\lam}$,
where $C(V)^{\lam}=\sum\limits_{\mu_1+\mu_2+\mu_3=\lam}V^{\mu_1}
\* \F^{\neu}(\chi)^{\mu_2}\* \F(L\sg_{>0})^{\mu_3}$.
Note that
we also have
\begin{align}\label{eq:bh-weights-of-C(V)}
C(V)=\bigoplus_{\mu\leq \lam}C(V)^{\mu}
 \text{ and }\dim_{\C} C(V)^{\mu}<\infty\
 \text{ for all $\mu\in \dual{\bh}$}
\end{align}
with an object $V$ of $\BGG_k^{\leq \lam}$.

\subsection{}
Define the odd operator
$d$
on $ C(V)$
by
\begin{align}
\begin{aligned}
  d\teigi \sum_\ud{\alpha\in \sroots_{>0}}{
n\in \Z}&(-1)^{p(\alpha)}
(u_{\alpha}(-n)+\Phi_{u_{\alpha}}(-n))\psi^{\alpha}(n)\\
&
-\frac{1}{2}\sum_\ud{\alpha,\beta,\gamma\in \sroots_{>0}}{
k+l+m=0}(-1)^{p(\alpha)p(\gamma)}([u_{\alpha},u_{\beta}]|u_{-\gamma})
\psi^{\alpha}(k)\psi^{\beta}(l)\psi_{\gamma}(m).
\end{aligned}\end{align}
Then,
we have
\begin{align}\label{eq:d2=0}
 d^2=0,\quad
d C^i(V)\subset C^{i+1}(V).
\end{align}
Thus, 
$(C(V),d)$ is a cohomology complex.
Define
\begin{align}
 H^{i}(V)\teigi H^i(C(V),d) \text{ with $i\in \Z$.}
\end{align}
\begin{Rem}\label{Rem;this-is-semi-inf-coh}
By  definition, we have
\begin{align*}
 \text{$
 H^{\bullet}(V)=H^{\semiinf+ \bullet}(L\sg_{>0},
V \* \F^{\neu}(\chi))
$,
}
\end{align*}
where 
$H^{\semiinf + \bullet}(L\sg_{>0},V)$
is the 
semi-infinite cohomology \cite{Feigin} of the Lie superalgebra
$L\sg_{>0}$ with coefficients in $V$.
\end{Rem}

\subsection{}
Decompose $d$
as $d=d^{\chi}+d^{\st}$,
where
\begin{align}
 d^{\chi}
&\teigi \sum_\ud{\alpha\in \sroots_{\frac{1}{2}}}{
n\geq 0}(-1)^{p(\alpha)}
\Phi_{\alpha}(n)\psi^{\alpha}(-n)
+\sum_{\alpha\in \sroots_{1}}(-1)^{p(\alpha)}
\chi(u_{\alpha}(-1))\psi^{\alpha}(1)
\end{align}
and $d^{\st}\teigi d-d^{\chi}$.
Then,
 by Lemma \ref{Lem:strange-def-of-weights},
we have
\begin{align}\label{eq:shift-of-weight-by-d}
 &d^{\chi} C(V)^{\lam}\subset
\sum_\ud{\alpha\in \sroots_1
}{ \bar{\chi}(u_{\alpha})\ne 0}C(V)^{\lam-\alpha+\delta}
 ,
\quad d^{\st}C(V)^{\lam}\subset C(V)^{\lam}
\end{align}
for all $\lam$.
Therefore,
by \eqref{eq:d2=0},
it follows that
\begin{align}
 (d^{\chi})^2=(d^{\st})^2=\{d^{\chi},d^{\st}\}=0.
\end{align}
\begin{Rem}
 We have
\begin{align}
 H^{\bullet}(C(V),d^{\st} )=H^{\semiinf +\bullet}(L\sg_{>0},
V\* \F^{\neu}(\chi_0)),
\end{align}
where $\F^{\neu}(\chi_0)$
is the $L\sg_{>0}$-module associated with the 
trivial character $\chi_0$ of $L\sg_{\geq 1}$
defined similarly  to  $\F^{\neu}(\chi)$.
\end{Rem}\subsection{}
Define 
\begin{align}
 \DW\teigi x+\D\in \bh,
\end{align}
Here, $x$ is the semisimple element in the $\sl_2$-triple,
as in Subsection 2.1.
Set
\begin{align}
 \bt\teigi \sh^f\+ \C \DW \subset \bh.
\end{align}
Let $\dual{\bt}$ be the dual of $\bt$.
For $\lam\in \dual{\bh}$,
let
$\xi_{\lam}\in \dual{\bt}$
denote its restriction to $\bt$.

\subsection{}
Let $V\in \Obj\BGG_k$
and let
\begin{align}
 C(V)=\bigoplus_{\xi\in \dual{\bt}}C(V)_{\xi},
\quad C(V)_{\xi}=\sum_\ud{\lam\in \dual{\bh}}{
\xi_{\lam}=\xi}C(V)^{\lam}
\end{align}
be the weight space  decomposition with respect to the
action of $\bt\subset
 \bh$.
Here and throughout,
we set
\begin{align*}
 M_{\xi}\teigi \{m\in M\mid t m=\bra \xi,t\ket m\
(\forall t\in \bt) \}
\end{align*}for a $\t$-module $M$.
By  \eqref{eq:shift-of-weight-by-d},
we see that
\begin{align*}
 \text{$d C(V)_{\xi}\subset C(V)_{\xi}$
for any $\xi\in \dual{\bt}$. 
}
\end{align*}
Hence,
the cohomology space
$H^{\bullet}(V)$ decomposes as
\begin{align}
 H^{\bullet}(V)=\bigoplus_{\xi\in \dual{\bt}}H^{\bullet}(V)_{\xi},
\quad H^{\bullet}(V)_{\xi}=H^{\bullet}(C(V)_{\xi},d).
\end{align}
Note
that
 the weight space
$C(V)_{\xi}$,
with $\xi\in \dual{\bt}$, is not finite dimensional in general
because $[\bt,e(-1)]=0$.
\begin{Rem}
 As discussed in Remark \ref{q:universal-Casimir},
the operator $\DW$
is essentially $-L(0)$,
where  $L(0)$ is the zero-mode 
of the Virasoro field, provided that $k\ne -h\che$.
\end{Rem}
\section{The assumption $f=f_{\theta}$}
\subsection{}
The gradation \eqref{eq:fun-gradation}
is called {\em minimal}
if 
\begin{align}
 \label{eq:minimal}
\text{$\sg=\sg_{-1}\+ \sg_{-\frac{1}{2}}\+\sg_0
\+\sg_{\frac{1}{2}}\+\sg_1$,\quad
$\sg_{-1}=\C f$
and 
\quad $\sg_1=\C e$.}
\end{align}
As shown in Section 5 of Ref.\ \cite{KW2003},
in this case,
one can choose a root system of $\sg$
so that
$e=e_{\theta}$
and $f=f_{\theta}$,
which are
the roots vectors attached to 
$\theta$
and $-\theta$,
where
$\theta$
is the corresponding highest root.

The condition \eqref{eq:minimal}
simplifies the theory in many ways.
{\em In this section we assume 
that  $f=f_{\theta}$
and
the condition \eqref{eq:minimal}
is satisfied.
}
Also,
we normalize $(~|~)$
as $(\theta|\theta)=2$.
\subsection{}
From the $\sl_2$-representation theory, we have
\begin{align}
 &\text{$\sg^f=\sg_{-1}\+ \sg_{-\frac{1}{2}}\+
\sg_0^f$
},\\
&\sg_0^f=\sn_{0,-}\+\sh^f\+\sn_{0,+}.
\end{align}
In particular,
\begin{align}
\sh=\sh^f \+ \C x,\quad 
 \sn_-\subset \sg^f,
\end{align}
and we have the exact sequence
\begin{align}\label{eq:10-28.exact-bt-bh}
 \begin{array}{ccccccccc}
  0&\rightarrow  &\C \alpha_0\+ \C \Lam_0&\hookrightarrow& \dual{\bh}&
\rightarrow &\dual{\bt}&\rightarrow &0\\
&&& &\lam &\mapsto &\xi_{\lam}.
 \end{array}
\end{align}
Here,
$\alpha_0=\delta-\theta$.
Therefore,
for $\lam$, $\mu\in \dual{\bh}_k$,
\begin{align}\label{eq:wight-equial}
 \text{ $\xi_{\lam}=\xi_{\mu}$
if and only if $\lam\equiv \mu\pmod{\alpha_0}$.
}
\end{align}
Let
$ \QW_+\subset \dual{\bt}$
be the image of $\bQ_+\subset \dual{\bh}$
in $\dual{\bt}$.
Then, by \eqref{eq:10-28.exact-bt-bh} we have
\begin{align}
 \bra \eta, \DW \ket \geq 0\quad \text{for all }\eta\in \QW_+.
\end{align}
Define a partial ordering
on $\dual{\bt}$
by $\xi\leq \xi'$
$\iff$
$\xi'-\xi\in \QW_+$.
Then,
for $\lam$, $\mu\in \dual{\bh}$,
we have the property 
\begin{align}\label{eq:map-of-paritaly}
 \text{$\xi_{\lam}\leq \xi_{\mu}$
 if $\lam\leq \mu$.
}
\end{align}
In particular,
\begin{align}\label{eq:dist-weight-theta}
 & V=\bigoplus_{\xi\leq \xi_{\lam}}V_{\xi}
\quad \text{for an object $V$ of $\BGG_k^{\leq \lam}$}.
\end{align}

\subsection{}
Let 
$\bg=\bigoplus_{\eta\in \dual{\bt}}(\bg)_{\eta}$
be the weight space decomposition  with respect to the
adjoint action of $\bt$.
Then,
 we have
\begin{align}\label{eq:bg-0}
 (\bg)_0=\bh\+ \C e(-1) \+ \C f(1)
\end{align}
(recall that $e=e_{\theta}$ and $f=f_{\theta}$).

Let $V\in \BGG_k$.
Then, each weight space
$V_{\xi}$,
where 
$\xi\in \dual{\bt}$,
is a module over $(\bg)_0$.
Also, 
we have
\begin{align}\label{eq:10-29-t-weight-spaces}
 \text{$V_{\xi_{\lam}}=\sum\limits_{\mu\in \dual{\bh}_k
\atop \mu\equiv \lam\pmod{\alpha_0}}V^{\mu}$
for $\lam\in \dual{\bh}_k$.}
\end{align}
Let
$(\sl_2)_0\cong \sl_2$
denote the subalgebra
of $(\bg)_0$
generated by
 $e(-1)$ and $f(1)$.

\subsection{}
Let $\BGG(\sl_2)$ 
be the Bernstein-Gel'fand-Gel'fand  category \cite{BGG}
of the Lie algebra $\sl_2=\bra e,x,f\ket$
(defined by  the commutation relations \eqref{def-of-sl2}).
That is,
$\BGG(\sl_2)$
is
the full subcategory of the category of left
$\sl_2$-modules 
consisting of modules $V$
such that 
(1) $V$ is finitely generated over $\sl_2$,
(2) $e$ is locally nilpotent on $V$,
(3) $x$ acts semisimply on $V$.

Let $\dBGG_k$
be the full subcategory of $\BGG_k$
consisting of objects $V$
such that
 each $V_{\xi}$,
with
$\xi\in \dual{\bt}$,
belongs to $\BGG(\sl_2)$ (viewed as a module over $(\sl_2)_0\cong \sl_2$).
It is clear that
$\dBGG_k$ is abelian.  
\begin{Lem}\label{Lem:dot-BGG}
$ $

 \begin{enumerate}
\item Any Verma module $M(\lam)$,
with
$\lam\in \dual{\bh}_k$,
belongs to $\dBGG_k$.
\item Any simple module $L(\lam)$,
with $\lam\in \dual{\bh}_k$,
belongs to $\dBGG_k$.
  \item Any object of $\BGG_k^{\tri}$ belongs to $\dBGG_k$.
  \item Any object of $\BGG_k^{\bigtriangledown}$ belongs to $\dBGG_k$.
 \end{enumerate}
\end{Lem}
\begin{proof}
(1) 
Certainly,
on $M(\lam)$,
$f(1)$ is locally nilpotent and
$h_0=[f(1),e(-1)]$ acts semisimply.
We have to show that
each $M(\lam)_{\xi}$,
with $\xi\in \dual{\bt}$,
is finitely generated over $(\sl_2)_0$. 
%
Clearly, this follows from
\eqref{eq:wight-equial},
\eqref{eq:map-of-paritaly} and the PBW theorem.
(2), (3) These assertions follow from the first assertion.
(4)
The category $\BGG(\sl_2)$
is closed under the operation of taking (graded) dual.
Hence, $\dBGG_k$ is closed under the operation of taking (graded) dual.
Therefore (4) follows from the third assertion.
\end{proof}
Let $\dBGG_k^{\leq \lam}$ be 
the full subcategory of $\BGG_k$ consisting of objects 
that belong to both $\dBGG_k$ and $\BGG_k^{\leq \lam}$.
Then, by Lemma \ref{Lem:dot-BGG},
$P_{\leq \lam}(\mu)$ and $I_{\leq \lam}(\mu)$
$(\mu\leq \lam)$ belong to $\dBGG_k^{\leq \lam}$.

The following proposition asserts  that
every object of $\BGG_k$
can be obtained as an injective limit of objects
of $\dBGG_k$.
\begin{Pro}\label{Pro:injective_limits}
 Let $V$ be an object of $\BGG_k$.
Then,
 there exits a sequence
$V_1\subset V_2\subset V_3\dots $
of objects of $\dBGG_k$
such that $V=\bigcup_i V_i$.
\end{Pro}
\begin{proof}
Note that,
because  each projective module $P_{\leq \lam}(\mu)$ 
 belongs to $\dBGG_k$,
so too do
finitely generated objects.
Let $\{0\}=V_0\subset 
V_1\subset V_2\subset V_3\dots $ be a highest weight filtration of
$V$,
so that
$V=\bigcup_i V_i$,
and each successive subquotient  $V_i/V_{i-1}$ is a
 highest weight module.
In particular,
each $V_i$ is finitely generated,
and hence
 belongs to $\dBGG_k$.
\end{proof}


\begin{Lem}
 \label{Lem:finitely-gen}
Let 
 $\xi \in \dual{\bt}$.
Then,
for any 
object $V$ of $\dBGG_k^{\leq \lam}$,
with $\lam\in \dual{\bh}_k$,
there exits a 
finitely generated submodule $V'$ of $V$
such that $(V/V')_{\xi'}=\{0\}$ if $\xi'\geq  \xi$,
where $\xi'\in \dual{\bt}$.
\end{Lem}
\begin{proof}
Let
$\P=\{v_1,v_2,\dots\}$
be a set of generators of $V$
such that 
(1) each 
$v_i$ belongs to $V^{\mu_i}$
for some $\mu_i\in \dual{\bh}$,
and
(2) if we set
$V_i=\sum_{r=1}^i U(\bg)v_r$ (and $V_0=\{0\}$),
then each  successive subquotient 
$V_{i}/V_{i-1}$ is a (nonzero) highest weight module with highest weight
 $\mu_i$.
(Therefore, $V_1\subset V_2\subset \dots $ is a highest weight filtration
of $V$.)
Then,
 by definition,
we have  $\sharp\{j\geq 1\mid \mu_j=\mu\}\leq 
[V: L(\mu)]$ for $\mu\in \dual{\bh}$.
Next, let
$
 \P_{\geq \xi}=\{ v_j\in \P\mid \xi_{\mu_j}\geq \xi\}\subset \P
$.
Then,
 by 
the definition of $\dBGG_k$,
$\P_{\geq \xi}$
is a finite subset of $\P$.
 The assertion follows,
because
$V'=\sum_{v\in \P_{\geq \xi}}U(\bg)v\subset V$
satisfies the desired properties.
\end{proof}
\subsection{}
Observe that we have the following:
\begin{align}
 & \F^{\neu}(\chi)=\bigoplus_{\xi\leq 0}\F^{\neu}(\chi)_{\xi},
\quad 
\dim_{\C} \F^{\neu}(\chi)_{\xi}<\infty\
(\forall \xi),\quad
\F^{\neu}(\chi)_0=\C \1_{\chi},\label{eq;f-d-F(chi)}
\\
& \F(L\sg_{>0})=\bigoplus_{\xi\leq 0}
\F(L\sg_{>0})_{\xi},
\quad \F(L(\sg_{>0}))_{0}
=\C\1\+ \C \psi_{\theta}(-1)\1\label{eq:F_0}
\end{align}
Thus,
if $V=\bigoplus_{\xi'\leq \xi}V_{\xi'}$ for some $\xi'\in \dual{\bt}$,
then $C(V)=\bigoplus_{\xi'\leq \xi}C(V)_{\xi'}$.
Hence we have the following assertion.
\begin{Lem}
\label{Lem:09-wt-appearing}
 Let
$V$ be an object of $\BGG_k$.
Suppose that $V=\bigoplus\limits_{\xi'\in \dual{\bt}
\atop \xi'\leq \xi}V_{\xi'}$ for some $\xi\in \dual{\bt}$.
Then,
$H^{\bullet}(V)=\bigoplus\limits_{\xi'\leq \xi}H^{\bullet}(V)_{\xi'}$.
In particular,
 $H^{\bullet}(V)=\bigoplus\limits_{\xi\leq \xi_{\lam}}H^{\bullet}(V)_{\xi}$
for $V\in \Obj\BGG^{\leq \lam}_k$.
\end{Lem}
\begin{Lem}
 \label{Lem:replace-by-finitely-gen}
Let $\xi \in \dual{\bt}$.
Then, for any 
object $V$ of $\dBGG_k^{\leq \lam}$,
where $\lam\in \dual{\bh}_k$,
there exits a 
finitely generated submodule $V'$ of $V$
such that $H^{\bullet}(V)_{\xi}=H^{\bullet}(V')_{\xi}$.
\end{Lem}
\begin{proof}
Let $V'$ be as in Lemma \ref{Lem:finitely-gen}.
Then, from  the exact sequence $0\rightarrow V'\rightarrow V\rightarrow
 V/V'
\rightarrow 0$,
we obtain the long exact sequence
\begin{align}\label{eq:10-28-exact-mornibg}
 \dots \rightarrow H^{i-1}(V/V')\rightarrow H^i(V')
\rightarrow H^i(V)\rightarrow H^i(V/V')\rightarrow \dots.
\end{align}
Clearly,
 the restriction of \eqref{eq:10-28-exact-mornibg}
 to the weight
 space
$\xi$ remains exact.
The desired result then follows from
Lemma \ref{Lem:finitely-gen}
and Lemma \ref{Lem:09-wt-appearing}.
\end{proof}
\subsection{}\label{sec:cohomology-of-highest-weight}
Here and throughout, 
we identify $\F(L(\sg_{>0}))_{0}$
with
the exterior power module
$\Lam (\C e(-1))$
by identifying $\psi_{\theta}(-1)$ with $e(-1)$
(see \eqref{eq:F_0}).
Let $\C_{\chi}$
be the one-dimensional module over
 the commutative Lie algebra
$\C e(-1)$
defined by the character
$\chi_{|\C e(-1)}$,
that is,
the
one-dimensional
$\C e(-1)$-module on which $e(-1)$
acts as the identity operator.
Also, let $V$ be an object of $\BGG_k$.
Then,
 $V_{\xi}\* \C_{\chi}$,
where $\xi\in \dual{\bt}$,
 is a module over $\C e(-1)$
by the tensor product action.
\begin{Lem}\label{Lem:cohomology-of-zero-mode}
Let $V \in \Obj\BGG^{\leq \lam}_k$,
with
$\lam\in \dual{\bh}_k$.
Then, we have
 \begin{align*}
  \text{$H^{i}(V)_{\xi_{\lam}}=
\begin{cases}
 H_{-i}(\C e(-1), V_{\xi_{\lam}}\* \C_{\chi})&(i=0,-1),\\
\{0\}&(\text{otherwise}).
\end{cases}
$
}
 \end{align*}
\end{Lem}
\begin{proof}
Because $V$ is an object of 
$\BGG_k^{\leq \lam}$,
we have
\begin{align*}
 C(V)_{\xi_{\lam}}
=V_{\xi_{\lam}}\* \F(L(\sg_{>0}))_{0}(=V_{\xi_{\lam}}\* \Lam (\C e(-1))).
\end{align*}
Next, observe that
\begin{align*}
d_{|C(V)_{\xi_{\lam}}}= \bar{d}=e(-1)\psi^{\theta}(1)+\psi^{\theta}(1).
\end{align*}
 From this,
 it follows that,
for $V\in \Obj\BGG_k^{\leq \lam}$,
the subcomplex
$(C(V)_{\xi_{\lam}},d)$
is identically the Chevalley
complex for calculating the (usual)
Lie algebra homology
$H_{\bullet}(\C e(-1),V_{{\xi}_{\lam}}\* \C_{\chi})$
(with the opposite grading).
\end{proof}
\subsection{}
Recall $\sl_2=\bra e,x,f\ket$.
Let 
$\bar{M}_{\sl_2}(a)\in \Obj\BGG (\sl_2)$
 be the Verma module of $\sl_2$ with highest weight 
$a\in \C$
and 
$\bar{L}_{\sl_2}(a)$ be
its unique simple quotient.
Here,
the highest weight is the largest eigenvalue of $2x$
(see  \eqref{def-of-sl2}).

Let 
$\C_{\bar{\chi}_-}$ be the one-dimensional
$\C f$-module on which $f$ acts as the identity operator.
\begin{Pro}
 \label{Pro:vanihsing-sl2}
$ $

\begin{enumerate}
\item
For  $a\in \C$,
 $H_i(\C f, \bar{M}_{\sl_2}(a)\* \C_{\bar{\chi}_-}
)=
\begin{cases}
 \C&(i=0),\\
\{0\}&(i=1).
\end{cases}$
\item
For $a\in \C$,
 $H_i(\C f, \bar{L}_{\sl_2}(a)\* \C_{\bar{\chi}_-}
)=
\begin{cases}
 \C&(i=0\text{ and }a\not\in \{0,1,2,\dots\}),\\
\{0\}&(\text{otherwise}).
\end{cases}$
\item
For  $a\in \C$,
 $H_i(\C f, \bar{M}_{\sl_2}(a)^*\* \C_{\bar{\chi}_-}
)=
\begin{cases}
 \C&(i=0),\\
\{0\}&(i=1).
\end{cases}$
 \item For
any object $V$ of $\BGG(\sl_2)$,
we have $H_1(\C f,V\* \C_{\bar{\chi}_-})=\{0\}$.
\item 
For
any object $V$ of $\BGG(\sl_2)$,
we have $\dim_{\C}H_0(\C f,V\* \C_{\bar{\chi}_-})<\infty$.
\end{enumerate}\end{Pro}
\begin{proof}
(1)
Since $\bar{M}_{\sl_2}(a)$ is free over $\C f$,
so is $\bar{M}_{\sl_2}(a)\* \C_{\bar{\chi}_-}$.
(2)
The case
in which
$a\not\in \{0,1,2,\dots\}$
follows from
the first assertion.
Otherwise,
$\bar{L}_{\sl_2}(a)$
is finite dimensional.
Hence,
$f$
is  nilpotent on $\bar{L}_{\sl_2}(a)$.
But this implies that
the corresponding Chevalley complex
is acyclic,
by the argument of  Theorem 2.3 of Ref.\ \cite{FKW}.
(3)
The case
in which
$a\not\in \{0,1,2,\dots\}$
follows from
the first assertion.
Otherwise,
we have the following exact sequence in $\BGG(\sl_2)$:
\begin{align*}
 0\rightarrow \bar{L}_{\sl_2}(a)\rightarrow
\bar{M}_{\sl_2}(a)^*
\rightarrow \bar{M}_{\sl_2}(-a-2)
\rightarrow 0.
\end{align*}
This induces the
 exact sequence
\begin{align*}
    0\rightarrow \bar{L}_{\sl_2}(a)\* \C_{\bar{\chi}_-}
\rightarrow
\bar{M}_{\sl_2}(a)^* \* \C_{\bar{\chi}_-}
\rightarrow \bar{M}_{\sl_2}(-a-2)\* \C_{\bar{\chi}_-}
\rightarrow 0.
  \end{align*}
Hence,
the assertion is obtained 
from the first and the second assertions by considering
the corresponding long exact sequence
of the Lie algebra homology.
(4)
Recall that
$\BGG(\sl_2)$ has enough injectives,
and each injective object $I$ admits a finite filtration such
that each successive quotient is isomorphic
to $\bar{M}_{\sl_2}(a)^*$ for some $a\in \C$.
Therefore,
the third assertion
implies that
$H_1(\C f, I\* \C_{\bar{\chi}_-})
=\{0\}$
for any injective object $I$ in $\BGG(\sl_2)$
(cf.\ Theorem 8.2 of \cite{A1}). 
For a given $V\in \Obj\BGG(\sl_2)$,
let
 $0\rightarrow V\rightarrow I\rightarrow V/I
\rightarrow 0$
be an exact sequence in $\BGG(\sl_2)$
such that $I$ is injective.
Then,
from 
the associated long exact
sequence,
it is proved that
$H_1(\C f, V\* \C_{\bar{\chi}_-})=\{0\}$.
(4) 
By the third assertion
the correspondence $V\mapsto H_0(\C f, V\* \C_{\bar{\chi}_-})$
defines an exact functor from $\BGG(\sl_2)$ to the category of
 $\C$-vector
spaces.
Because
the assertion follows from the first assertion  for Verma modules,
 it also holds for any projective object $P$ of 
$\BGG(\sl_2)$, since $P$ has a (finite) Verma flag.
This completes the proof, as  $\BGG(\sl_2)$ has enough projectives.
\end{proof}
\subsection{}
For $\lam\in \dual{\bh}$,
let 
\begin{align*}
 |\lam\ket =v_{\lam}\* \1\in C(M(\lam)).
\end{align*}
Then,
 $d |\lam\ket =0$,
and thus $|\lam\ket $ defines an element of $H^0(M(\lam))$.
Again with a slight abuse of notation,
we denote
the image of $|\lam\ket$ under the natural map
$C(M(\lam))\rightarrow C(L(\lam))$
by
$|\lam\ket$.
Also,
let
\begin{align}\label{eq:caninical-vector-dual}
  |\lam\ket ^*=v_{\lam}^*\* \1\in H^0(M(\lam)^*),
\end{align}
where $v_{\lam}^*$ is the vector of $M(\lam)^*$
dual to $v_{\lam}$.
\begin{Pro}\label{Pro:theVanishin-zeromode-1}
For any $\lam\in \dual{\bh}$,
we have the following:
\begin{enumerate}
\item $H^i(M(\lam))_{\xi_{\lam}}=
\begin{cases}
 \C |\lam\ket&(\text{if $i=0$}),\\
\{0\}&(\text{otherwise}).
\end{cases}$
 \item $H^i(L(\lam))_{\xi_{\lam}}=
\begin{cases}
 \C |\lam\ket&(\text{if $i=0$ and $\bra \lam,\alpha_0\che\ket\not\in
\{0,1,2,\dots\}$}),\\
\{0\}&(\text{otherwise}).
\end{cases}$
\item $H^i(M(\lam)^*)_{\xi_{\lam}}=\begin{cases}
		  \C |\lam\ket^*&(\text{if $i=0$}),\\
\{0\}&(\text{otherwise}).
		 \end{cases}$
\end{enumerate}
\end{Pro}
\begin{proof}
Observe that
$M(\lam)_{\xi_{\lam}}$
is isomorphic to
$\bar{M}_{\sl_2}(\bra \lam,\alpha_0\che\ket)$
as a module
over
$(\sl_2)_0$.
Similarly,
$L(\lam)_{\xi_{\lam}}$ and $M(\lam)_{\xi_{\lam}}^*$
are isomorphic to $\bar{L}_{\sl_2}(\bra \lam,\alpha_0\che\ket)$
and $\bar{M}_{\sl_2}(\bra \lam,\alpha_0\che\ket)^*$, respectively.
Hence,
the assertion
  follows from Lemma \ref{Lem:cohomology-of-zero-mode}
and
Proposition \ref{Pro:vanihsing-sl2}.
\end{proof}
\subsection{}
Let $V$ be an object of $\dBGG_k$.
For
a given
$\xi\in \dual{\bt}$,
consider the (usual) Lie algebra
homology
$H_{\bullet}(\C e(-1),V_{\xi}\* \C_{\chi})$.
It is calculated using
the Chevalley complex 
$(V_{\xi}\* \Lam(\C e(-1)),\bar{d})$,
(see Section \ref{sec:cohomology-of-highest-weight}).
The action of $\bt$
on $V_{\xi}$ commutes with $\bar{d}$.
Thus,
there is a natural action of $\bt$
on $H_{\bullet}(\C e(-1),V\* \C_{\chi})$;
\begin{align*}
 H_{\bullet}(\C e(-1),V\* \C_{\chi})=
\bigoplus_{\xi\in \dual{\bt}}H_{\bullet}(\C e(-1),V\* \C_{\chi})_{\xi}.
\end{align*}
By definition, we have  
\begin{align}\label{eq:10-28-by-definition}
 H_{\bullet}(\C e(-1),V\* \C_{\chi})_{\xi}
=H_{\bullet}(\C e(-1),V_{\xi}\* \C_{\chi})
\quad \text{for }\xi \in \dual{\bt}.
\end{align}
Hence,
from 
\eqref{eq:dist-weight-theta}
and
Proposition \ref{Pro:vanihsing-sl2},
we obtain the following assertion.
\begin{Pro}\label{Pro:vanishing-and-fd-e-1}
 Let $V$
be   an object of $\dBGG_k$.
Then,
we have the following:
\begin{enumerate}
 \item $H_1(\C e(-1),V\* \C_{\chi})=\{0\}$.
\item $H_0(\C e(-1),V\* \C_{\chi})=\bigoplus_{\xi\leq \xi_{\lam}}
H_0(\C e(-1),V\* \C_{\chi})_{\xi}$
if $V\in \Obj\dBGG_k^{\leq \lam}$
with $\lam\in \dual{\bh}$.
\item
Each weight space
$H_0(\C e(-1),V\* \C_{\chi})_{\xi}$,
$\xi \in \dual{\bt}$,
is finite dimensional.
\end{enumerate}
\end{Pro}
\subsection{}
We end this section with the following important proposition.
\begin{Pro}
\label{Pro:estimate}
For any
object $V$
of $\dBGG_k$,
each weight space 
$H^{\bullet}(V)_{\xi}$,
where $\xi \in \dual{\bt}$, 
is finite dimensional.
Moreover,
if $V\in \Obj \dBGG_k^{\leq \lam}$,
with $\lam\in \dual{\bh}$,
then
$H^i(V)_{\xi}=\{0\}$
if $
\frac{1}{2}|i|>
\bra \xi_{\lam}-\xi,\DW\ket
$
 for $i\in \Z$.
\end{Pro}
\begin{proof}
We may assume that $V\in \Obj\dBGG_k^{\leq \lam}$
for some $\lam\in \dual{\bh}_k$.
Decompose $\F(L\sg_{>0})$ as 
$\F(L\sg_{>0})=\F(L\sg_{>0}/\C e(-1))\* \Lam (\C e(-1))$,
where 
$\F(L(\sg_{>0})/\C e(-1))$
is the subspace of $\F(L(\sg_{>0}))$
spanned by the vectors
\begin{align*}
 \psi_{\alpha_1}(m_1)\dots
\psi_{\alpha_r}(m_r)
\psi^{\beta_1}(n_1)\dots
\psi^{\beta_s}(n_s)\1,
\end{align*}
with $\alpha_i,\beta_i\in \sroots_{>0}$,
$m_i\leq \begin{cases}
	-2&(\text{if }\alpha_i=\theta),
\\-1&(\text{otherwise}), 
	\end{cases}$
$n_i\leq 0$.
Then, we have
\begin{align*}
 \text{$\F^n(L\sg_{>0})=\sum_{i-j=n}\F^i(L\sg_{>0}/\C e(-1))
\* \Lam^j (\C
 e(-1))$,
}
\end{align*}where 
$\F^i(L\sg_{>0}/\C e(-1))=\F(L\sg_{>0}/\C e(-1))\cap \F^i(L\sg_{>0})$.

Now, define
\begin{align}
 G^p C^n(V)\teigi V\* \F^{\neu}(\chi)\* 
\sum\limits_\ud{i-j=n}{ i\geq p}\F^i(L\sg_{>0}/\C e(-1))\* \Lam^j (\C
 e(-1))\subset C^n(V).
\end{align}
Then,
we have
\begin{align*}
&C^n(V)=G^n C^n(V)
\supset  G^{n+1} C^n(V)\supset G^{n+2}C^n(V)=\{0\},\\
&d G^p C^n(V)\subset G^p C^{n+1}(V).
\end{align*}
The corresponding
the spectral sequence,
$E_r\Rightarrow H^{\bullet}(V)$,
is
the
 {\em Hochschild-Serre spectral sequence}
(more precisely, the semi-infinite,  Lie superalgebra analogue of
this spectral sequence)
for the ideal $\C e(-1)\subset L\sg_{>0}$.
By  definition,
we have the isomorphism
\begin{align*}
 E_1^{p,q}
&=H_{-q}(\C e(-1), V\* \F^{\neu}(\chi))\* \F^p(L\sg_{>0}/\C e(-1)),
\end{align*}
because the complex 
$(\sum_p G^{p} C(V)/G^{p+1}C(V),d)$
is identical to the corresponding Chevalley complex.
By Proposition \ref{Pro:vanishing-and-fd-e-1} (1),
we have
\begin{align}\label{eq:Pro-estimate-E_1}
 E_1^{p,q}
\cong \begin{cases}
H_0(\C e(-1), V\* \C_{{\chi}})\* 
\F^{\neu}(\chi)\* \F^p(L\sg_{>0}/
\C e(-1))
&(q=0)\\
		  \{0\}&(q\ne 0)
		 \end{cases}
\end{align}
as $\bt$-modules
for any $p$.

Next, observe that
\begin{align*}
&\F^p(L\sg_{>0}/
\C e(-1))=\bigoplus_{\xi\leq 0
}
\F^p(L\sg_{>0}/
\C e(-1))_{\xi},\\
&
\dim_{\C}\F^p(L\sg_{>0}/
\C e(-1))_{\xi}<\infty \ 
\text{ for any } \xi,\\
&
\F^p(L\sg_{>0}/
\C e(-1))_{\xi}=\{0\}
\text{ unless }
\bra \xi,\DW\ket\leq -\frac{1}{2}|p|.
\end{align*}
Hence,
from
\eqref{eq;f-d-F(chi)},
Proposition \ref{Pro:vanishing-and-fd-e-1} 
and \eqref{eq:Pro-estimate-E_1},
it follows that
\begin{align}
  E_1^{p,0}
=
\bigoplus_\ud{\xi\leq \xi_{\lam}
}{ \frac{1}{2}|p|\leq
\bra \xi_{\lam}-\xi,\DW\ket}
(E_1^{p,0})_{\xi},\quad
\dim_{\C} (E_1^{p,0})_{\xi}<\infty \
\text{ for any } \xi.
\end{align}
The assertion is thus   proved,
as our filtration
is compatible with the action of $\bt$.
\end{proof}
\section{The Kac-Roan-Wakimoto construction II: the $\W$-algebra
construction of superconformal algebras}
In this section 
 we recall the definition of the $\W$-algebra
$\W_k(\sg,f)$ 
and collect necessary information about its structure.

\subsection{}
Let
$V_k(\g)=U(\bg)\*_{U(\sg\*\C[t]\+\C K\+ \C \D)}
\C_k\in \Obj\BGG_k$
be the universal affine vertex algebra associated with $\sg$ at 
a given level
$k\in \C$.
Here,
$\C_k$ is the one-dimensional representation  of 
$\sg\*\C[t]\+\C K\+ \C \D$ on which
$\sg\*\C[t]\+ \C \D$ acts trivially and $K$ acts as $k\id$.
Hence,
 $V_k(\sg)$ is a quotient of $M(k \Lam_0)$
as a $\bg$-module.
It is known that the space 
$V_k(\sg)$ has a natural vertex (super)algebra structure,
and
the space
\begin{align}
 C(V_k(\g))=V_k(\g)\* \F^{\neu}(\chi)\* \F(L\sg_{>0})
\end{align}
also has  a natural vertex (super)algebra structure
(see Ref.\ \cite{KRW} for details).
Let
$|0\ket=(1\*1)\*\1_{\chi}\* \1$ be the canonical vector.
Also, let $Y(v,z)\in \End C(V_k(\g))[[z,z\inv]]$
be the field corresponding to $v\in C(V_k(\sg))$.
Then,
by the definition,
\begin{align*}
 &Y(v(-1)|0\ket,z)=v(z)
= \sum_{n\in \Z}v(n)z^{-n-1}\quad
\text{for }v\in \sg,\\
&
Y(\Phi_{\alpha}(-1)|0\ket,z)=\Phi_{\alpha}(z)=
\sum_{n\in \Z}\Phi_{\alpha}(n)z^{-n-1}\quad
\text{for }\alpha\in \sroots_{\frac{1}{2}},
\\
&
 Y(\psi_{\alpha}(-1)|0\ket,z)= \psi_{\alpha}(z)=
\sum_{n\in \Z}\psi_{\alpha}(n)z^{-n-1}
\quad 
\text{for }\alpha\in \sroots_{>0},\\
& 
Y(\psi_{-\alpha}(0)|0\ket,z)=\psi_{-\alpha}(z)=
\sum_{n\in \Z}\psi_{-\alpha}(n)z^{-n}
\quad
\text{for } \alpha\in \sroots_{>0}.
\end{align*}
\subsection{}
Define
\begin{align}\label{eq:Def-of-W-algebras}
\W_k(\sg,f)\teigi H^0(V_{k}(\g)).
\end{align}
Then,
$Y$ descends to the map
\begin{align}\label{eq:field_of_W}
Y:\W_{k}(\sg,f)\rightarrow
\End \W_{k}(\sg,f) [[z,z\inv]],
\end{align}
because,
as shown in Ref.\ \cite{KRW},
the following relation holds:
\begin{align}\label{eq:commutativiry_of_d_with_Y}
 [d,Y(v,z)]=Y(d v,z)\quad\text{for all $v\in C(V_k(\g))$.}
\end{align}
Therefore,
$\W_{k}(\sg,f)$
has a vertex algebra  structure.
The vertex  algebra  $\W_{k}(\sg,f)$
is called the
{\em $\W$-$($super$)$algebra associated with the pair $(\sg,f)$
at level $k$}.
By definition,
the vertex algebra  $\W_{k}(\sg,f)$
acts naturally 
on $H^i(V)$,
where $V\in \BGG_k$,
$i\in \Z$.
Thus,
we obtain a  family
of  functors
$V\fmap H^i(V)$,
depending on $i\in \Z$,
from $\BGG_k$
to the category of $\W_k(\sg,f)$-modules.
\begin{Rem}$ $

\begin{enumerate}
 \item  If $\sg$ is a Lie algebra and $f$ is a regular nilpotent
element of $\sg$,
then
$\W_k(\sg,f)$ is identical to $\W_k(\g)$,
the $\W$-algebra defined by B. L.  Fe\u{\i}gin and E. V.  Frenkel \cite{FF_W}.
\item  V. G. Kac, S.-S. Roan and M. Wakimoto 
gave a more general definition of $\W$-algebras
(see Ref.\ \cite{KRW} for details).
\end{enumerate}
\end{Rem}
\subsection{}
As shown in Ref.\  \cite{KRW},
the vertex algebra
$\W_k(\sg,f)$ has a superconformal algebra structure
provided that
the level
$k$ is non-critical,
i.e.,
that
$k+h\che\ne 0$.
Here,
 $h\che$
is the dual Coxeter number of $\sg$.
Let $L(z)=\sum_{n\in \Z}L(n)z^{-n-2}$
be the corresponding Virasoro field.
The explicit form of $L(z)$
is given in Ref.\ \cite{KRW}.
If
$f=f_{\theta}$,
 its central charge is given by 
\begin{align}
 c(k)=\frac{k \sdim \sg}{k+h\che}-6k+h\che-4.
\end{align}
Let
\begin{align}
 S(z)=\sum_{n\in \Z}S(n)z^{-n-2}\teigi 2(k+h\che)L(z).
\end{align}
Then,
$S(z)$ is well-defined for any level $k$.
\begin{Rem}\label{q:universal-Casimir}
Let
$\widehat{\Omega}$ be the universal Casimir operator \cite{KacBook}
of $\bg$
acting on
$V\in \BGG_k$.
Then,
we have
\begin{align*}
S(0)+ 2(k+h\che)\DW
=\widehat{\Omega}
\end{align*}
on $H^{\bullet}(V)$.
\end{Rem}
\subsection{}
Let 
\begin{align*}
 J^{(v)}(z)=
\sum_{n\in \Z}J^{(v)}(n)z^{-n-1}=v(z)+
\sum_{\beta,\gamma\in \roots_{>0}}
(-1)^{p(\gamma)}([v,u_{\beta}]|u_{-\gamma}):\psi_{\gamma}(z)\psi^{\beta}(z):,
\end{align*}
for $v\in \sg_{\leq 0}$.

Also, let
$C_k(\g)$
be the subspace of $C(V_k(\sg))$
spanned by the vectors
\begin{align*}
 J^{(u_1)}(m_1)\dots J^{(u_p)}(m_p)
\Phi_{\alpha_1}(n_1)\dots \Phi_{\alpha_q}(n_q)
\psi^{\beta_1}(s_1)\dots \dots \psi^{\beta_r}(s_r)|0\ket
\end{align*}
with $u_i\in \sg_{\leq 0}$,
$\alpha_i\in \sroots_{\frac{1}{2}}$,
$\beta_i\in \sroots_{>0}$,
$m_i,n_i,s_i\in \Z$.
As shown in Ref.\ \cite{KW2003},
$C_k(\g)$ is a vertex subalgebra
and a subcomplex of $C(V_k(\g))$.
Moreover,
it was proved that
\begin{align}\label{eq:realization-of-W-algebras}
 \W_k(\sg,f)=H^0(C_k(\g),d)
\end{align}
as vertex algebras.
This follows from the tensor product decomposition of
the complex $C(V_k(\sg))$ (see Ref.\ \cite{KW2003} for details).
\subsection{}
Let
\begin{align*}
& \bgf\teigi 
\sgf\* \C[t,t\inv]\+\C 1
\end{align*}
be the affine Lie superalgebra of $\sgf$
with respect to the
$2$-cocycle
$(~,~)^\nat$,
defined
by 
\begin{align}\label{eq:2-cocycle}
 (u\*t^m,v\* t^n)^\nat=
\begin{cases}
 m\delta_{m,n}
\left((k+h\che)(u|v)-\frac{1}{2}\str_{\sg_0}
(\ad u) (\ad v)\right)& (
\text{if }u,v\in \sg_0),\\
0&\text{(otherwise).}
\end{cases}\end{align}
Also,
let
$V_k^{\nat}(\sgf)$ 
be the corresponding 
universal vertex affine algebra:
\begin{align}
 V_{k}^{\nat}(\sgf)\teigi U(\bgf)\*_{U(\sgf\* \C[t]\+ \C 1)}\C.
\end{align}
Then,
the correspondence
\begin{align*}
 v\* t^n\rightarrow J^{(v)}(n)\quad \text{for }v\* t^n\in \bgf
\end{align*}
defines a  $V_k^\nat(\sgf)$-module structure 
on $C(V)$,
$V\in \BGG_k$. 
In particular,
we have the following embedding
of vertex algebras:
\begin{align}\label{eq:emb-alg}
 V_k^{\nat}(\sgf)\hookrightarrow C_k(\g)\subset C(V_k(\g)).
\end{align}

\begin{Th}
[{V. G. Kac and M. Wakimoto: Theorem 4.1 of Ref.\ \cite{KW2003}}]\label{th:str-of-W}
There exists a filtration
\begin{align*}
&\{0\}= F_{-1} \W_k(\sg,f)\subset F_0 \W_k(\sg,f)
\subset F_1 \W_k(\sg,f)\subset \dots \nonumber \\
\end{align*}
 of $\W_k(\sg,f)=H^0(C_k(\g),d)$
such that
\begin{align}
 &\W_k(\sg,f)=\bigcup_p F_p \W_k(\sg,f),\nonumber \\
&  \bt \cdot F_p \W_k(\sg,f)\subset F_p \W_k(\sg,f)\quad \text{for all }p,
\nonumber \\
 &F_p\W_k(\sg,f)\cdot  F_q\W_k(\sg,f)
\subset F_{p+q}\W_k(\sg,f) \quad \text{for all }p,\ q,
\label{eq:henna-kigou}
\end{align}
where the left-hand-side of \eqref{eq:henna-kigou}
denotes the span of the vectors ,
\begin{align*}
 Y_n(v)w\quad (v\in F_p \W_k(\sg,f),\ w\in F_q \W_k(\sg,f), n\in \Z),
\end{align*}
and such that
the map \eqref{eq:emb-alg}
induces the isomorphism
\begin{align*}
 V_k^\nat(\sgf)\cong 
\gr_F \W_k(\sg,f)=\bigoplus F_p \W_k(\sg,f)/F_{p-1}\W_k(\sg,f)
\end{align*}
as vertex algebras and $\bt$-modules.
\end{Th}
\begin{Rem}
 Actually,
  stronger results were proved by
V. G. Kac and M. Wakimoto  \cite{KW2003}.
Specifically,
it was shown that
$H^i(V_k(\g))= H^i(C_k(\g),d)=\{0\}$
$(i\ne 0)$.
Their proof is based on the argument given in Ref.\  \cite{FB}.
Further,
the explicit form of $\W_k(\sg,f)$
was obtained for the case
$f=f_{\theta}$.
\end{Rem}

For $v\in \sg^f$,
let $W^{(v)}\in \W_k(\sg,f)$
be the cocycle in $C_k(\sg)$ corresponding to $v(-1)|0\ket 
\in V_k^{\natural}(\sg^f)$.
Let us write
\begin{align}
 Y(W^{(v)},z)=\sum_{n\in \Z}W^{(v)}(n)z^{-n-1}.
\end{align}
Then,
 by Theorem \ref{th:str-of-W},
the following map defines an isomorphism of $\bt$-modules:
\begin{align*}
 \begin{array}{ccc}
 U(\sgf\* \C[t\inv]t\inv)&\rightarrow & \W_k(\sg,f), \\
u_1(n_1)\dots u_r(n_r)&\mapsto
& W^{(u_1)}(n_1)\dots W^{u_r}(n_r)|0\ket.
 \end{array}
\end{align*}
We remark that  the filtration $\{F_p\W_k(\sg,f)\}$
in Theorem \ref{th:str-of-W}
is now described
by
\begin{align}\label{eq:description-of-filtration}
\begin{aligned}
 & F_p \W_k(\sg,f)\\
&=
\haru\{ W^{(u_1)}(n_1)\dots W^{(u_r)}(n_r)|0\ket\mid 
r\geq 0,\ n_i\in \Z, \ u_i\in \sgf_{-s_i},\
\sum_{i=1}^r s_i\leq  p\}.
\end{aligned}\end{align}
\begin{Rem}\label{Rem:wieght-no-ugokasikata}
$ $

\begin{enumerate}
 \item 
Let $V\in \BGG_k$.
Then,
we have
\begin{align*}
 W^{(v)}(n)H^{\bullet}(V)_{\xi}\subset
H^{\bullet}(V)_{\xi+\eta}
\quad (\text{if }v(n)\in (\bgf)_{\eta}).
\end{align*}
Thus, in particular,
\begin{align*}
 [\DW, W^{(v)}(n)]=(n-j)W^{(v)}(n)\quad 
\text{if }v\in \sgf_{-j}
\end{align*}
in $\End (H^{\bullet}(V))$.
\item  $W^{(f)}(n)$ coincides with $S(n-1)$
 up to a nonzero multiplicative factor.
\end{enumerate}
\end{Rem}
\section{
 Irreducible highest weight representations and their characters
}
\label{section:Irr-HW}
{\em
We assume 
that  
$f=f_{\theta}$
and
the condition \eqref{eq:minimal}
is satisfied
for the remainder of the paper.
}
\subsection{}
Decompose
$\bgf$
as 
\begin{align}
 \bgf=(\bgf)_-\+ \hW\+ (\bgf)_+,
\end{align}
where
\begin{align*}
&\left(\bgf\right)_+
\teigi \C f\* \C[t]t^2
\+(\sg_{-\frac{1}{2}}\+ \sn_{0,-}\+ \sh^f)\*\C[t]t
\+ \sn_{0,+}\* \C[t],\\
&\ \hW
\teigi \sh^f\+ \C 1\+ \C f\* \C t,
\\
& \left(\bgf\right)_-\teigi (\C f\+ \sg_{-\frac{1}{2}}\+
\sn_{0,-})\* \C[t\inv]\+ (\sh^f\+ \sn_{0,+})\* \C[t\inv]t\inv.
\end{align*}
By definition,
 $\hW$ is commutative.
Let $V$ be a $\W_{k}(\sg,f)$-module.
Then,
the correspondence
\begin{align}\label{eq:rep-of-cartan-w}
 h\mapsto W^{(h)}(0)\ (h\in \sh^f), \quad f\* t^1\mapsto S(0)
\end{align}
defines the action of $\hW$ on $V$ (see Ref.\ \cite{KRW}).
Below we shall regard a $\W_k(\sg,f)$-module as a
module over
$\hW$ via the correspondence \eqref{eq:rep-of-cartan-w}.
Note that, for $k\ne -h\che$,
the $\hW$-action on $H^{\bullet}(V)$,
with $V\in \Obj \BGG_k$,
is essentially  the same as the $\bt$-action
on it.

The following lemma is obvious 
(see Remark \ref{Rem:wieght-no-ugokasikata}).
\begin{Lem}
 The operator $W^{(u)}(n)$,
with $u(n)\in (\bgf)_+$,
is  locally nilpotent on $H^{\bullet}(V)$,
for $V\in \Obj \BGG_k$.
\end{Lem}
\subsection{}
A  $\W_k(\sg,f)$-module
$V$
is called {\em $\QW_+$-gradable}
 if  
it admits a decomposition 
$V=\bigoplus_{\xi\in \QW_+}V[-\xi]$
such that 
$W^{(v)}(n) V[-\xi]\subset V[-\xi+\eta]$
for $v(n)\in (\bgf)_{\eta}$,
$ \xi\in \QW_+$ and  $\forall \eta\in \QW$.
A $\QW_+$-gradable $\W_k(\sg,f)$-module 
$V$ is called a  
{\em highest weight module}
with {\em highest weight}
$\phi\in\dual{\hW}$
if there exists a non-zero vector $v_{\phi}$
(called a {\em highest weight vector})
such that
\begin{align*}
 & \text{$V$ is generated by $v_{\phi}$ over $\W_k(\sg,f)$},\\
&W^{(u)}(n)v_{\phi}=0
\quad \text{(if $u\* t^n\in (\bgf)_+$),}\\
&W^{(h)}(0)v_{\phi}={\phi}(h)v_{\phi}\quad \text{(if $h\in \sh^f$),}
\\
&S(0)v_{\phi}=\phi(f\* t) v_{\phi}.
\end{align*}

Let
$B=\{b_j\mid j\in J\}$
be a PBW 
basis of
$U((\bgf)_-)$
of the form
\begin{align*}
 b_j=
(u_{j_1}\*t^{n_{j_1}})\dots
(u_{j_r}\*t^{n_{j_r}}).
\end{align*}Then,
a highest weight module
is spanned by the vectors
\begin{align*}
 W^{(b_j)}\teigi W^{(u_{j_1})}(n_{j_1})\dots W^{(u_{j_r})}(n_{j_r})v_{\phi}.
\end{align*}
A highest weight module $V$ with highest weight vector 
$v_{\phi}$ is called a {\em Verma module}
if 
the above vectors
$W^{(b)}$,
with $b\in B$,
forms a 
basis of $V$ (see Ref.\ \cite{KW2003}).
Let 
$\M(\phi)$
denote
the Verma module with highest weight $\phi\in \dual{\hW}$.
\begin{Rem}
 Let
\begin{align*}
& F_p \M(\phi )\\
&=
\haru\{ W^{(u_1)}(n_1)\dots W^{(u_r)}(n_r)v_{\phi}\mid 
r\geq 0,\ n_i\in \Z,\ u_i\in \sgf_{-s_i},\
\sum_{i=1}^r s_i\leq  p\}.
\end{align*}
Then,
$\{F_p \M(\phi)\}$
defines an increasing
filtration of $\M(\phi)$
such that
the corresponding graded space
$\gr_F \M(\phi)$
is naturally a module  over $\gr_F \W_k(\sg,f)$
and isomorphic to $U(\bgf)\*_{U(\hW\+ (\bgf)_+
)}\C_{\phi}$,
where $\C_{\phi}$
is a $\hW
\+ (\bgf)_+$-module on which $(\bgf)_+$
acts trivially 
and $\hW$ acts through the character $\phi$.
\end{Rem}

\subsection{}
For $\lam\in \dual{\bh}_k$,
define $\phi_{\lam}\in \dual{\hW}$
by 
\begin{align}\label{eq:def-of-map-W-hw}
 \begin{aligned}
 & \phi_{\lam}
(h)=\lam(h)\ \text{for }h\in \sh^f,\\
&
\phi_{\lam}(f\* t)=
|\lam+\rho|^2-|\rho|^2-
2(k+h\che)\bra\lam,\DW\ket
 \end{aligned}\end{align}
(cf.\ Remark \ref{q:universal-Casimir}).
Here,
$\rho=\srho+h\che\Lam_0$,
and $\srho$ is equal to the half of
the difference of the sum of positive 
even roots and the sum of positive odd roots of $\sg.$
Then,
the correspondence $\dual{\bh}_k\ni\lam\mapsto \phi_{\lam}
\in \dual{\hW}$
is surjective.

By 
Proposition \ref{Pro:theVanishin-zeromode-1} (1),
there is a natural homomorphism of $\W_k(\sg,f)$-modules
of the form
\begin{align}\label{eq:map-Verma}
\begin{array}{ccc}
  \M(\phi_{\lam})&\rightarrow& H^0(M(\lam)) \\
v_{\phi_{\lam}}&\mapsto & |\lam\ket.
\end{array}
\end{align}
\begin{Th}[V. G. Kac and M. Wakimoto:
Theorem 6.3 of Ref.\ {\cite{KW2003}}]\label{Th:image-of-Verma}
 For each $\lam\in \dual{\bh}_k$, with $k\in \C$,
we have
$H^i(M(\lam))=\{0\}$
for $i\ne 0$,
and the map \eqref{eq:map-Verma} is an isomorphism
of $\W_k(\sg,f)$-modules:
$H^0(M(\lam))\cong \M(\phi_{\lam})$.
\end{Th}
\subsection{}\label{subsection:re-def-of-L}
Let $\phi\in \dual{\hW}$.
Choose $\lam\in \dual{\bh}_{k}$
such that $\phi_{\lam}=\phi$.
Let
$\WN(\phi)$
be the 
the sum of all $\DW$-stable proper
 $\W_{k}(\sg,f)$-submodules of
$
H^0(M(\lam))=\M(\phi_{\lam})$ (Theorem \ref{Th:image-of-Verma}).
Then,
$\WN(\phi)\subset \M(\phi)$ is independent of the choice of $\lam$,
as long as
$\phi=\phi_{\lam}$ holds.
We define
\begin{align}
 \why(\phi)\teigi \M(\phi)/\WN(\phi)\quad \text{with }\phi\in \dual{\hW}.
\end{align}
It is clear that
in the case that $k\ne -h\che$,
$\why(\phi)$ is the unique irreducible quotient of $\M(\phi)$.
In the case that  $k=-h\che$,
the following assertion 
is proved together with  Theorem \ref{Th;image-of-simples}.
\begin{Th}\label{Th:irr-11-1}
The $\W_k(\sg,f)$-module
 $\why(\phi)$,
with $\phi \in \dual{\hW}$, is irreducible. 
\end{Th}
It is clear that the set $\{ \why(\phi)\mid \phi \in \dual{\hW}\}$
represents the complete set of the isomorphism classes of irreducible
highest weight modules over $\W_{k}(\sg,f)$.



\subsection{}

\begin{Th}
\label{Th:Vanishing-Projective}
For any object $V$ in $\Obj \BGG_k^{\tri}$,
 we have $H^i(V)=\{0\}$  $(i\ne 0)$.
In particular,
we have $H^i(P_{\leq \lam}(\mu))=\{0\}$ $(i\ne 0)$
for any $\lam,\mu\in \dual{\bh}_k$
such that $\mu\leq \lam$.
\end{Th}
\begin{proof}
By  Theorem \ref{Th:image-of-Verma},
 the assertion can be shown by  induction applied to
 the length of 
a Verma flag of $V\in \Obj\BGG_{k}^{\tri}$
(cf. Theorem 8.1 of Ref.\ \cite{A1}).
\end{proof}
\begin{Th}
 \label{Th:vanishing-positive}
For any object $V$ of $\BGG_k$
we have $H^i(V)=\{0\}$ for all $i>0$.
\end{Th}
\begin{proof}
 We may assume that
$V\in \Obj \BGG_k^{\leq \lam}$ for some $\lam\in \dual{\bh}_k$.
Also, by Proposition \ref{Pro:injective_limits},
 we may assume that
$V\in \Obj \dBGG_k^{\leq \lam}$, 
since the cohomology functor commutes with
injective limits.
Clearly,
 it is sufficient to show that $H_i(V)_{\xi}=\{0\}$ $(i>0)$
for each $\xi\in \dual{\bt}$.
By Lemma \ref{Lem:replace-by-finitely-gen},
for a given $\xi$,
there exists a finitely generated submodule 
$V'$ of $V$ such that
\begin{align}
 H^{i}(V)_{\xi}=H^{i}(V')_{\xi}\quad \text{for all }i\in \Z.
\label{eq;in-the-proof-yale1}
\end{align}
Because $V'$ is finitely generated,
 there exists a 
projective module $P$
of the form $\bigoplus_{i=1}^r P_{\leq \lam}(\mu_i)$
and an exact sequence
$
 0\rightarrow N\rightarrow P\rightarrow V\rightarrow 0
$
in $\dBGG^{\leq \lam}_k$.
Considering the corresponding long exact sequence,
 we obtain
\begin{align}
 \dots \rightarrow H^i(P)\rightarrow H^i(V)\rightarrow H^{i+1}(N)
\rightarrow H^{i+1}(P)
\rightarrow \dots.
\end{align}
Hence, it follows that
$H^i(V')\cong H^{i+1}(N)$ for all $i>0$,
by Proposition \ref{Th:Vanishing-Projective}.
Therefore,
we find
\begin{align}
 H^i(V)_{\xi}\cong  H^{i+1}(N)_{\xi} \quad \text{for all }i>0,
\end{align}
by \eqref{eq;in-the-proof-yale1}.
Then,
because $N\in \Obj\dBGG_k^{\leq \lam}$,
we can repeat this argument to find,
for each $k>0$,
some object $N_k$ of $\dBGG_k^{\leq \lam}$
such that \begin{align}
 H^i(V)_{\xi}\cong  H^{i+k}(N_k)_{\xi} \quad \text{for all }i>0.
\end{align}
But 
by Proposition \ref{Pro:estimate},
this implies that
$H^i(V)_{\xi}=\{0\}$
for $i>0$.
\end{proof}
\subsection{}
\begin{Th}
\label{Th;vanishing-and-cocyclic-dual-2}
 For any $\lam\in \dual{\bh}_k$,
we have
 $H^i(M(\lam)^*)=\{0\}$
for $i\ne 0$.
\end{Th}
Theorem \ref{Th;vanishing-and-cocyclic-dual-2}
is proved in Subsection \ref{subsection:proof-of-vanishing-of-dual-of-Verma}.
\begin{Th}
\label{Th:cofree-new}
  Suppose that
$\bra \lam,\alpha_0\che\ket\not\in \{0,1,2,\dots\}$.
Then, any nonzero $\W_k(\sg,f)$-submodule of 
$H^0(M(\lam)^*)$ 
contains the canonical vector $|\lam\ket ^*$
of $H^0(M(\lam)^*)$.
\end{Th}

Theorem
    \ref{Th:cofree-new}
is proved 
together with 
 Theorem \ref{Th:verma-in-cofree-proof}.
\begin{Rem}
 Theorem \ref{Th:cofree-new} holds without any restriction on $\lam$.
Indeed,
it  can seen from Corollary \ref{Co:exactness}
and (the proof of) Theorem \ref{Th;image-of-simples}
that
if $\bra \lam,\alpha_0\che\ket \in
\{0,1,2,\dots\}$,
then
$H^0(M(\lam)^*)\cong H^0(M(r_0\circ \lam)^*)$,
where $r_0$ is the reflection corresponding to $\alpha_0$.
\end{Rem}
The following theorem
is
a
consequence of
Theorem \ref{Th;vanishing-and-cocyclic-dual-2}  
which can be proved in the same manner
as Theorem 8.1 of Ref.\ \cite{A1}.
 \begin{Th}
\label{Th:vanihing-injective}
  For a given $\lam\in \dual{\bh}_k$,
$H^i(I_{\leq \lam}(\mu))=\{0\}$
$(i\ne 0)$
for any $\lam,\mu\in \dual{\bh}_k$
such that $\mu\leq \lam$.
 \end{Th}
 Using Theorem \ref{Th:vanihing-injective},
the following assertion can be  proved in the same manner as
Theorem \ref{Th:Vanishing-Projective}.
\begin{Th}
\label{Th:vanishing-negative}
For any object $V$ of $\BGG_k$
we have $H^i(V)=\{0\}$ for all $i<0$.
\end{Th}
\subsection{Main results}
From  Theorems \ref{Th:vanishing-positive} and
  \ref{Th:vanishing-negative},
we obtain the following results.
\begin{Th}
\label{Th:vanishin}
Let $k$ be an arbitrary complex number.
We have
$H^i(V)=\{0\}$
$(i\ne 0)$
for any object $V$ in $\BGG_k$.
\end{Th}
\begin{Rem}\label{Rem:vanihing-conj}
By Theorem \ref{Th:vanishin},
we have, in particular,
$H^i(L(\lam))=\{0\}$ $(i\ne 0)$
for each $\lam\in \dual{\bh}$.
This was conjectured by V. G. Kac, S.-S. Roan and M. Wakimoto \cite{KRW}
in  the case that $\lam$ is admissible.
\end{Rem}

\begin{Co}
 \label{Co:exactness}
For any $k\in \C$, the correspondence $V\fmap H^0(V)$
defines an exact functor from $\BGG_k$
to the category of $\W_k(\sg,f)$-modules.
\end{Co}

\begin{Th}
\label{Th;image-of-simples}
Let $k$ be an arbitrary complex number
and
let $\lam\in \dual{\bh}_k$.
If 
 $\bra \lam,\alpha_0\che\ket \in
\{0,1,2,\dots\}$,
then
$  H^0(L(\lam))=\{0\}$.
Otherwise
$H^0(L(\lam))$ is an 
irreducible $\W_{k}(\sg,f)$-module
that is isomorphic to $\why(\phi_{\lam})$.
\end{Th}
\begin{proof}
Let $N(\lam)$ be the unique maximal proper submodule of $M(\lam)$.
Then, we have an exact sequence
$0\rightarrow N(\lam)\rightarrow M(\lam)\rightarrow L(\lam)\rightarrow
 0$ in $\BGG_k$.
With this,
using Corollary \ref{Co:exactness},
we obtain
an exact sequence
\begin{align}\label{eq:09-exact-proof-irr}
 0\rightarrow H^0(N(\lam))\rightarrow
H^0(M(\lam))
\overset{\pi}{\rightarrow}
 H^0(L(\lam))\rightarrow 0.
\end{align}
By Theorem \ref{Th:image-of-Verma},
we have
$H^0(M(\lam))\cong \M(\phi_{\lam})$.
Therefore
\eqref{eq:09-exact-proof-irr} shows that
$H^0(L(\lam))$
is generated over $\W_k(\sg,f)$
by the image $\pi(|\lam\ket)$
of the highest weight vector $|\lam\ket\in H^0(M(\lam))$.
Thus
it follows that
$H^0(L(\lam))=\{0\}$
if and only if 
$\pi(|\lam\ket)\ne \{0\}$.
But $\pi(|\lam\ket)\ne \{0\}$
if and only if
$H^0(L(\lam))_{\xi_{\lam}}\ne \{0\}$,
because 
$\dim_{\C} H^0(M(\lam))_{\xi_{\lam}}=1$,
by Proposition \ref{Pro:theVanishin-zeromode-1} (1).
Hence,
 by
Proposition \ref{Pro:theVanishin-zeromode-1} (2),
it follows  that
$H^0(L(\lam))$ is nonzero
if and only if $\bra \lam,\alpha_0\che\ket \not\in\{0,1,2,\dots\}$.

Next,
 suppose that
$\bra \lam,\alpha_0\che\ket \not\in\{0,1,2,\dots\}$,
so that $H^0(L(\lam))\ne \{0\}$.
Also, let $N$ be a nonzero  submodule of $H^0(L(\lam))$.
As
$L(\lam)$ is a submodule of $M(\lam)^*$,
$H^0(L(\lam))$  
is also  a submodule of $H^0(M(\lam)^*)$,
 by 
the exactness of the functor $H^0(?)$ (Corollary \ref{Co:exactness}).
Hence,
 $N$ is a submodule of $H^0(M(\lam)^*)$.
But then,
 by Theorem \ref{Th:cofree-new},
it follows that
$ \pi(|\lam\ket)\in N$.
Therefore $N$ must coincide with the entire space $H^0(L(\lam))$.
We have thus shown that
$H^0(L(\lam))$ is irreducible.
Finally,
we have to show (in the case $k=-h\che$) that
$\WN(\phi_{\lam})
=H^0(N(\lam))$ 
($\WN(\phi)$ is defined in Section \ref{subsection:re-def-of-L}.)
Clearly,
$H^0(N(\lam))$ is stable under the action of $\DW$.
Thus,
$\WN(\phi_{\lam})\supset H^0(N(\lam))$.
In addition, 
 $\WN(\phi_{\lam})\subset H^0(N(\lam))$,
 by the irreduciblity
of
$H^0(L(\lam))=H^0(M(\lam))/H^0(N(\lam))$,
which we have just proved.
This completes the proof.
\end{proof}
\begin{Rem}
Theorem \ref{Th;image-of-simples}
 was conjectured by V. G. Kac, S.-S. Roan, M. Wakimoto \cite{KRW}
in the case of an admissible weight $\lam$.
\end{Rem}

\subsection{The characters}
For an object $V$ of $\dBGG_k$,
define the formal characters
\begin{align*}
&\ch V\teigi \sum_{\lam\in \dual{\bh}}e^{\lam}\dim_{\C}V^{\lam},\\
&\ch H^{0}(V)\teigi \sum_{\xi\in \dual{\bt}}e^{\xi} \dim_{\C} H^0(V)_{\xi}.
\end{align*}
In addition,
for $\lam,\mu\in \dual{\bh}$
define an integer $[L(\lam):M(\mu)]$
by
\begin{align*}
 \ch L(\lam)=\sum_{\mu\in \dual{\bh}}[L(\lam): M(\mu)]\ch M(\mu).
\end{align*}
Then,
 by the exactness of the functor $H^0(?)$ (see Corollary \ref{Co:exactness}),
we have
\begin{align}\label{eq:ch-formula}
 \ch H^0(L(\lam))
=\sum_{\mu\in \dual{\bh}}[L(\lam): M(\mu)]\ch H^0(M(\mu)).
\end{align}
Because 
the correspondence $\dual{\bh}_k\ni \lam\mapsto \phi_{\lam}\in \dual{\hW}$
(see \eqref{eq:def-of-map-W-hw})
is
 surjective,
it follows from \eqref{eq:ch-formula},
Theorem \ref{Th:image-of-Verma}
 and Theorem
\ref{Th;image-of-simples} that
the character of {\em any} irreducible  highest weight 
representation
$\why(\phi)$
of $\W_k(\sg,f_{\theta})$
at {\em any} level $k\in \C$
is determined by the character of the corresponding 
$\bg$-module $L(\lam)$.

\begin{Rem} $ $

\begin{enumerate}
 \item For an admissible weight $\lam$,
the character formula \eqref{eq:ch-formula}
was
conjectured by V. G. Kac, S.-S. Roan, M. Wakimoto \cite{KRW}.
\item 
By Theorem \ref{Th:vanishin},
it follows that
\eqref{eq:ch-formula}
  can be  obtained by 
calculating the 
Euler-Poincar\'{e} character of $H^0(L(\lam))$
(see Ref.\ \cite{KRW} for details).
\item In the case that $\sg$ is a Lie algebra
and $k\ne -h\che$,
the number $[L(\lam):M(\mu)]$
is known.
(It can be  expressed in terms of the Kazhdan-Lusztig polynomials.
The most general formula is given in  Ref.\ \cite{KT}).
\item In the case that  $\sg={\mathfrak{spo}}(2|1)$,
 $\W_k(\sg,f_{\theta})$
is the Neveu-Schwarz algebra.
All minimal series representations
of the Neveu-Schwarz algebra (see, e.g., Ref.\ \cite{KW1})
can be obtained from the admissible
${\mathfrak{spo}}(2|1\widehat{)}$-modules \cite{KW1},
as explained  by V. G. Kac, S.-S. Roan and
M. Wakimoto  \cite{KRW}.
\item In the case that  $\sg={\mathfrak{sl}}(2|1)$,
 $\W_k(\sg,f_{\theta})$ is the $N=2$ superconformal algebra.
The minimal series representations of
the $N=2$ superconformal algebra
(cf.\ Refs.\ \cite{D, Ki, M})
can be obtained from the admissible
${\mathfrak{sl}}(2|1\widehat{)}$-modules \cite{KW1},
as explained
 by V. G. Kac, S.-S. Roan and
M. Wakimoto  \cite{KRW}.
\item For further examples and references,
see Refs.\ \cite{KRW, KW2003, KW2004-2}.
\end{enumerate}
\end{Rem}

\section{The computation of $H^{\bullet}(M(\lam)^*)$}
\label{Sec:cruciela}
In this section,
we prove Theorems \ref{Th;vanishing-and-cocyclic-dual-2}
and  \ref{Th:cofree-new}.
Specifically,
we compute
$H^{\bullet}(M(\lam)^*)$,
with $\lam\in \dual{\bh}_k$.
This is done by
using a spectral sequence
which
 we  define in 
Subsection \ref{subsection;def-of-ss}.
It is a version of  the Hochschild-Serre spectral sequence
for the subalgebra
$\C e(-1)\+ \sg_{>0}\* \C[t]\subset L\sg_{>0}$.
\subsection{}
As in
\eqref{eq:def-of-N-chi-+},
let
\begin{align}
 N(\chi_-)\teigi U(L\sg_{<0})/U(L\sg_{<0})\ker\chi_-.
\end{align}
Here,
$\ker\chi_-\subset U(L\sg_{\leq -1})$
is the kernel 
of the character
$\chi_-$ of $L\sg_{\leq -1}$,
defined by
\begin{align}
 \chi_-(u(m))\teigi (e(-1)|u(m)),
\text{ where }u\in \sg_{\leq -1}\text{ and } m\in \Z.
\end{align}
Let $\Phi_u(n)$,
with
$u\in \sg_{<0} $ and $ n\in \Z$,
denote the image of $u(n)\in L\sg_{<0}$
in $N(\chi_-)$.
As above,
we set
$\Phi_{-\alpha}(n)=\Phi_{u_{-\alpha}}(n)$
with $\alpha\in \sroots_{\frac{1}{2}}$ and $ n\in \Z $.
Then,
the correspondence
$\Phi_{\alpha}(n)\mapsto \Phi_{-\alpha}(-n)$,
with
$\alpha\in \sroots_{\frac{1}{2}}$ and $ n\in \Z$,
defines the anti-algebra isomorphism
\begin{align*}
N(\chi)\cong N(\chi_-).
\end{align*}

\subsection{}
Let
$\F^{\neu}(\chi_-)$ be the irreducible representation
of $N(\chi_-)$ generated by a vector $\1_{\chi_-}$
such that 
$\Phi_{-\alpha}(n)\1_{\chi_-}=0$
for $\alpha\in \sroots_{\frac{1}{2}}$
and $n\geq 1$.
As in the case considered above,
we define a semisimple action 
of $\bh$
on $\F^{\neu}(\chi_-)$
by
$h\1_{\chi_-}=0$,
$\Phi_{-\alpha}(n)\F^{\neu}(\chi_-)^{\lam}
\subset 
\F^{\neu}(\chi_-)^{\lam-\alpha+n\delta}$,
with
$h\in \sh$,
$\alpha\in \sroots_{\frac{1}{2}}$,
$n\leq 0$,
and $
\lam\in \dual{\bh}$.
Then,
$\F^{\neu}(\chi_-)
=\bigoplus_{\xi\in\dual{\bt}}
\F^{\neu}(\chi_-)_{\xi}$
and
$\dim \F^{\neu}(\chi_-)_{\xi}<\infty$
for all $\xi$.

\subsection{}
There exists a unique bilinear form
\begin{align}
 \text{$\bra\cdot |\cdot \ket^{\neu}:
  \F^{\neu}(\chi)\times  \F^{\neu}(\chi_-)\rightarrow \C $
}
\end{align}such that 
$\bra\1_{\chi}|\1_{\chi_-}\ket^{\neu}=1$
and $\bra\Phi_{\alpha}(m)v|v'\ket^{\neu}=
\bra v|\Phi_{-\alpha}(-m)v'\ket^{\neu}$,
where
$v\in \F^{\neu}(\chi)$,
$v'\in \F^{\neu}(\chi_-)$,
$\alpha\in \sroots_{\frac{1}{2}}$
and $m\in \Z$.
It is easy to see that
this form is non-degenerate.
Indeed,
its restriction 
on
$\F^{\neu}(\chi)_{\xi}\times
\F^{\neu}(\chi_-)_{\xi}$,
with $\xi \in \dual{\bt}$,
is 
non-degenerate. 
Hence, we have
\begin{align}\label{eq:identi-f-dual1}
 \F^{\neu}(\chi)=
\F^{\neu}(\chi_-)^*,
\end{align}
since
each space
$\F^{\neu}(\chi_-)_{\xi}$,
with $\xi\in \dual{\bt}$,
decomposes into a finite sum of finite-dimensional weight spaces
$\F^{\neu}(\chi_-)^{\lam}$.

\subsection{}
Let $\Cl(L\sg_{<0})$ be the Clifford superalgebra
associated with $L\sg_{<0}\+ \dual{(L\sg_{<0})}$
and its natural bilinear form.
It is generated by the elements
$\psi_{-\alpha}(n)$
and
${\psi}^{-\alpha}(n)$
with $\alpha\in \sroots_{<0}$ and $n\in \Z$
which satisfy the relations
$[\psi_{-\alpha}(m),\psi^{-\beta}(n)]=\delta_{\alpha,\beta}\delta_{m+n,0}$.
Here 
the parity of
$\psi_{-\alpha}(n)$ and
${\psi}^{-\alpha}(n)$
  is reverse to $u_{-\alpha}$.
We have
an anti-algebra isomorphism
$\Cl(L\sg_{>0})\cong \Cl(L\sg_{<0})$
defined by
$\psi_{\alpha}(m)\mapsto (-1)^{p(\alpha)}\psi_{-\alpha}(-m)$
and
$\psi^{\alpha}(m)\mapsto \psi^{-\alpha}(-m)$,
where $\alpha\in \sroots_{>0}$ and $m\in \Z$.
\subsection{}
Let $\F(L\sg_{<0})$ be the irreducible representation
of $\Cl(L\sg_{<0})$
generated by the vector $\1_-$ with the properties
$\psi_{-\alpha}(n)\1_-=0$,
where $\alpha\in \sroots_{>0}$, $n\geq 1$,
and
$\psi^{-\alpha}(n)\1_-=0$,
where
$\alpha\in \sroots_{>0}$,
$n\geq 0$.
As above,
we have a natural action of $\bh$ on $\F(L\sg_{<0})$.

There exists a unique bilinear form
\begin{align}\label{eq:identi-f-dual2}
 \text{$\bra\cdot |\cdot \ket^{\ch}:
  \F(L\sg_{>0})\times  \F(L\sg_{<0})\rightarrow \C $,
}
\end{align}
which is non-degenerate 
on
$\F(L\sg_{>0})^{\lam}\times
\F(L\sg_{<0})^{\lam}$,
with $\lam\in \dual{\bh}$,
such that 
$\bra\1|\1_-\ket^{\ch}=1$,
 $\bra\psi_{\alpha}(n)v|v'\ket^{\ch}=
(-1)^{p(\alpha)}\bra v|\psi_{-\alpha}(-n)v'\ket^{\ch}$
and
 $\bra\psi^{\alpha}(n)v|v'\ket^{\ch}=
\bra v|\psi^{-\alpha}(-n)v'\ket^{\ch}$,
where
$v\in \F(L\sg_{>0})$,
$v'\in \F(L\sg_{<0})$,
$\alpha\in \sroots_{>0}$ and
$n\in \Z$.
Hence,
we have
\begin{align}
 \F(L\g_{>0})=\F(L\g_{<0})^*.
\end{align}
\subsection{}
Let
\begin{align*}
 \text{$C_-(V)=V\* \F^{\neu}(\chi_-)\* \F(L\sg_{<0})$}
\quad \text{with }V\in \Obj\BGG_k.
\end{align*}
Then,
$C_-(V)=\bigoplus_{\lam\in \dual{\bh}}C_-(V)^{\lam}$
with respect to the diagonal action of $\bh$.
By
\eqref{eq:identi-f-dual1}
and 
\eqref{eq:identi-f-dual2},
we
 have 
\begin{align}\label{eq:iden-comp-of-dual}
 C(\dual{V})=C_-(V)^*\quad \text{for }V\in \Obj \BGG_k
\end{align}
as $\C$-vector spaces.
Here again,
${}^*$ is defined by \eqref{eq:def-of-dual}.
Under the identification \eqref{eq:iden-comp-of-dual},
we have
\begin{align}
 (d g)(v)=g(d_- v)\quad (g\in C(V^*), v\in C(V) ),
\end{align}
where
\begin{align*}
\begin{aligned}
  d_-=\sum_\ud{\alpha\in \sroots_{>0}}{
n\in \Z}(-1)^{p(\alpha)}
&(u_{-\alpha}(-n)+\Phi_{u_{-\alpha}}(-n)){\psi}^{-\alpha}(n)\\
-\frac{1}{2}&\sum_\ud{\alpha,\beta.\gamma\in \sroots_{>0}}{
k+l+m=0}(-1)^{p(\alpha)p(\gamma)}
(u_{\gamma}|[u_{-\alpha},u_{-\beta}]) {\psi}^{-\alpha}(k)\psi^{-\beta}(l)
\psi_{-\gamma}(m).
\end{aligned}\end{align*}
Clearly, we have $d_-^2=0$.
Also,
$d_-$
decomposes as
\begin{align}\label{eq:dif-of-double-}
\begin{aligned}
 & d_-=d_-^{\chi}+d_-^{\st},\\
&(d_-^{\chi})^2=(d_-^{\st})^2=\{d_-^{\chi},d_-^{\st}\}=0,
\end{aligned}\end{align}
where
\begin{align}
 d_-^{\chi}=\sum_\ud{\alpha\in \sroots_{\frac{1}{2}}}{ n\geq 1}
(-1)^{p(\alpha)}\Phi_{-\alpha}(n)\psi^{-\alpha}(-n)
+
\sum_{\alpha\in \sroots_{1}}
(-1)^{p(\alpha)}\chi_-(u_{-\alpha}(1))\psi^{-\alpha}(-1)
\end{align}
and $d_-^{\st}=d_--d_-^{\chi}$.
\begin{Rem}
By Theorem 2.3 of \cite{FKW},
the complex
$(C_-(V),d_-)$ is acyclic 
for any $V\in \Obj \BGG_k$,
since  $f(1)$ is locally nilpotent on $V$
(see Remark \ref{Rem:remark-of-red-coh}).
\end{Rem}
\subsection{}
It is clear that
$C_-(V_k(\sg))$ possesses a natural vertex algebra structure.
The correspondences
$v(n)\mapsto v^t(-n)$,
$\psi_{\alpha}(n)\mapsto (-1)^{p(\alpha)}\psi_{-\alpha}(-n)$,
$\psi^{\alpha}(n)\mapsto \psi^{-\alpha}(-n)$,
$\Phi_{\alpha}(n)\mapsto \Phi_{-\alpha}(-n)$
extend to the
anti-algebra homomorphism
\begin{align}\label{eq:anti-algebra}
{}^t :\ \U(C(V_k(\g))\rightarrow \U(C_-(V_k(\g))),
\end{align}
where
$\U(C(V_k(\g)))$ and $\U(C_-(V_k(\g)))$
are universal enveloping algebras
of $C(V_k(\g))$ and $C_-(V_k(\g))$ respectively
in the sense of Ref.\ \cite{FZ}.
Note that we have the relations
$d_-=d^t$,
$d^{\st}_-=(d^{\st})^t$
and
$d^{\chi}_-=(d^{\chi})^t$.

\subsection{}
For $v\in \sg$, define
\begin{align*}
 J^{(v)}_-(z)
&=\sum_{n\in \Z}J^{(v)}_-(n)z^{-n-1}\\
&\teigi 
v(z)+\sum_{\alpha,\beta\in \sroots_{>0}}(-1)^{p(\gamma)}
(u_{\gamma}|[v,u_{-\beta}]):\psi_{-\gamma}(z)\psi^{-\beta}(z):,
\end{align*}
where
$\psi_{-\alpha}(z)=\sum_{n\in \Z}\psi_{-\alpha}(n)z^{-n}$
and
$\psi^{-\alpha}(z)=\sum_{n\in\Z}\psi^{-\alpha}(z)z^{-n-1}$
with $\alpha\in \sroots_{>0}$.
Then,
 we have
\begin{align}
 J^{(v)}(n)^t=J^{(v^t)}_-(-n)
\quad \text{for }v\in \sg^f,\ n\in \Z.
\end{align}
\subsection{}
Let $\lam\in \dual{\bh}_k$
and
let
$C_-(\lam)$
be the 
subspace of
$C_-(M(\lam))$
spanned by the vectors
\begin{align*}
 J^{(u_1)}_-(m_1)\dots J^{(u_p)}_-(m_p)
\Phi_{-\alpha_1}(n_1)\dots \Phi_{-\alpha_q}(n_q)
\psi^{-\beta_1}(s_1)\dots \dots \psi^{-\beta_r}(s_r)|\lam\ket_-,
\end{align*}
with $u_i\in \sg_{\geq 0}$,
$\alpha_i\in \sroots_{\frac{1}{2}}$,
$\beta_i\in \sroots_{>0}$
and
$m_i,n_i,s_i\in \Z$,
where 
$|\lam\ket_-$
is the canonical vector $v_{\lam}\* \1_{\chi_-}\* \1_-$.
Then,
$d_- C_-(\lam)\subset C_-(\lam)$,
i.e.,
$C_-(\lam)$ is a subcomplex of $C_-(M(\lam))$.


The inclusion $C_-(\lam)\hookrightarrow C_-(M(\lam))$
induces the surjection
\begin{align}\label{eq:fininal-surj-comp}
 C_-(M(\lam))^*\twoheadrightarrow C_-(\lam)^*.
\end{align}
Let the differential $d$
act on $C_-(\lam)^*$
as $(d g )(v)=g( d_- v)$
with $g\in C_-(\lam)^*$, $v\in C_-(\lam)$.
Then, the space $H^{i}(C_-(\lam)^*,d)$,
where $i\in \Z$,
 is naturally a module over $\W_k(\sg,f)=H^0(C_k(\sg))$,
and
 \eqref{eq:fininal-surj-comp}
induces the  homomorphism
\begin{align}\label{eq:homo-dual-W-10}
 H^{\bullet}(M(\lam)^*)\rightarrow H^{\bullet}(C_-(\lam)^*,d)
\end{align}
of $\W_k(\sg,f)$-modules.
The action of $\W_k(\sg,f)$ on $H^{i}(C_-(\lam)^*,d)$
is described by
\begin{align}\label{eq:action-of-W-transpose}
 (W^{(v)}(n)g)(v)=g(W^{(v^t)}(-n)v),
\end{align}
where $W^{(v^t)}_-(-n)$
is the image of $W^{(v)}(n)\in H^0(\U(C_k(\g)),\ad d) $
under the map \eqref{eq:anti-algebra}.

The following proposition can be shown
in the same manner as Proposition 6.3 of \cite{A1}.
\begin{Pro}\label{Pro:first-identification-dual}
For any $\lam\in \dual{\bh}$,
the map \eqref{eq:homo-dual-W-10} is an isomorphism
of $\W_k(\sg,f)$-modules and $\bt$-modules:
 \begin{align*}
 H^{\bullet}(M(\lam)^*)\cong H^{\bullet}(C_-(\lam)^{*},d).
\end{align*}
\end{Pro}
Below we compute the cohomology $H^{\bullet}(C_-(\lam)^*)=
H^{\bullet}(C_-(\lam)^{*},d)$.

\subsection{}
Here we employ
the notation of the previous subsections.
First, note that
\begin{align*}
 C_-(\lam)=\bigoplus_{\xi\leq \xi_{\lam}}C_-(\lam)_{\xi}.
\end{align*}
Also,
observe that
the subcomplex 
$C_-(\lam)_{\xi_{\lam}}
\subset C_-(\lam)$
is 
spanned by the vectors
\begin{align*}
 J^{(e_{\theta})}_-(-1)^n
|\lam\ket_-,\quad 
 J^{(e_{\theta})}_-(-1)^n
\psi^{-\theta}(-1)|\lam\ket_-
\end{align*}
with $n\in \Z_{\geq 0}$.
(Hence,
 $C_-(\lam)_{\xi_{\lam}}$ is infinite dimensional.)
 Let
\begin{align}\label{eq:def-of-Gp-bar}
 G_p C_-(\lam)_{\xi_{\lam}}
=\sum_\ud{\mu \in \dual{\bh}}{ -\bra \mu-\lam,x\ket\leq p}
C_-(\lam)_{\xi_{\lam}}^{\mu}\subset C_-(\lam)_{\xi_{\lam}}\quad
\text{for }p\leq 0.
\end{align}
Thus,
$ G_p C_-(\lam)_{\xi_{\lam}}$
is
a subspace of $C_-(\lam)_{\xi_{\lam}}$
spanned by the vectors
\begin{align*}
 J_-^{(e_{\theta})}(-1)^n
|\lam\ket_-,\quad 
 J_-^{(e_{\theta})}(-1)^{n-1}
\psi^{-\theta}(-1)|\lam\ket_-,\quad \text{with }n\geq -p.
\end{align*}
Now, the following assertion is clear.
\begin{Lem}
 The space $C_-(\lam)_{\xi_{\lam}}/G_p C_-(\lam)_{\xi_{\lam}}$
is finite dimensional for each $p\leq 0$.
\end{Lem}
Define 
$G_p C_-(\lam)$,
with
$p\leq 0$,
as the subspace of $C_-(\lam)$
spanned by the vectors
\begin{align*}
 J_-^{(u_1)}(m_1)\dots J_-^{(u_p)}(m_p)
\Phi_{-\alpha_1}(n_1)\dots \Phi_{-\alpha_q}(n_q)
\psi^{-\beta_1}(s_1)\dots \dots \psi^{-\beta_r}(s_r)v,
\end{align*}
with $u_i\in \sg_{\geq 0}$,
$\alpha_i\in \sroots_{\frac{1}{2}}$,
$\beta_i\in \sroots_{>0}$,
$m_i,n_i,s_i\in \Z$
and
$v\in  G_p C_-(\lam)_{\xi_{\lam}}$.
%
Then, we have
\begin{align}\label{eq:filt-C--(lam)}
\begin{aligned}
 & \dots \subset G_p C_-(\lam)
\subset G_{p+1} C_-(\lam)
\subset \dots \subset G_0 C_-(\lam)
=C_-(\lam),\\
&\bigcap_p G_p C_-(\lam)
=\{0\},\\
&d_- G_p C_-(\lam)
\subset G_{p}C_-(\lam).
\end{aligned}\end{align}

The following
Lemma is easily proven.
\begin{Lem}\label{Lem:GpC-is-submodule}
$ $

\begin{enumerate}
 \item  The subspace $G_p C_-(\lam)$, with $p\leq 0$,
is preserved under the
the action of the operators
$J_-^{(u)}(n)$ with $u\in \sg_{\geq 0},\ n\in \Z$,
$\Phi_{-\alpha}(n)$ with $\alpha\in \sroots_{\frac{1}{2}},\ n\in \Z$
and
$\psi^{-\alpha}(n)$ with $\alpha\in \sroots_{>0},\ n\in \Z$.
\item For each $p\leq 0$,
$C_-(\lam)/G_p C_-(\lam)$ is a direct sum of finite dimensional
weight spaces $(C_-(\lam)/G_pC_-(\lam))_{\xi}$, with $\xi\in \dual{\bt}$.
\end{enumerate}\end{Lem}
\subsection{}
\label{subsection;def-of-ss}
For a semisimple $\bt$-module $X$,
we define
\begin{align}
 D(X)\teigi \bigoplus_{\xi}\Hom_{\C}(X_{\xi},\C).
\end{align}
Next, let
\begin{align}
 G^p C_-(\lam)^*
\teigi \left(C_-(\lam)/G_p C_-(\lam)
\right)^*\subset 
C_-(\lam)^*\quad \text{for }p\leq 0,
\end{align}
where ${}^*$ is defined by \eqref{eq:def-of-dual}.
Then,
by Lemma \ref{Lem:GpC-is-submodule},
we have
\begin{align*}
 G^p C_-(\lam)^*
=D(C_-(\lam)/G_p C_-(\lam)),
\end{align*}
and  $G^p C_-(\lam)^*$ is a $C_k(\sg)$-submodule of $C_-(\lam)^*$.
Also, by \eqref{eq:filt-C--(lam)}, we have
\begin{align}
\begin{aligned}
 & \dots \supset G^p C_-(\lam)^*
\supset G^{p+1} C_-(\lam)^*
\supset \dots \supset G^0
 C_-(\lam)^*
 = \{0\},\\
&C_-(\lam)^*=\bigcup_p G^p C_-(\lam)^*,\\
&d G^p C_-(\lam)^*
\subset G^{p}C_-(\lam)^*.
\end{aligned}
\end{align}
Therefore we obtain
the corresponding  spectral sequence
$E_r\Rightarrow H^{\bullet}(C_-(\lam)^*)=H^{\bullet}(M(\lam)^*)$.
By  definition,
\begin{align}\label{eq:E-1-first-identity}
 E_1^{\bullet,q}=H^q(\gr^G C_-(\lam)^*
,d),
\end{align}
where
$\gr^G C_-(\lam)^*
=\sum_p G^p C_-(\lam)^*/G^{p+1}C_-(\lam)^*$.
Moreover, because
our filtration is compatible with the action of $\bt$,
each $E_r$ is a direct sum of
$\t$-weight spaces:
\begin{align*}
 E_r=\bigoplus_{\xi\in \dual{\bt}}(E_r)_{\xi}.
\end{align*}
In particular,
we have
\begin{align}\label{eq:convergenceof-sp-seqence-04}
( E_r)_{\xi}
\Rightarrow H^{\bullet}(C_-(\lam)^*)_{\xi}=H^{\bullet}(M(\lam)^*)_{\xi}
\quad \text{for each }\xi\in \dual{\bt}.
\end{align}
Below we compute this spectral sequence.

\subsection{}
Consider the exact sequence
\begin{align*}
 0\rightarrow G_{p+1}C_-(\lam)/G_{p}C_-(\lam)
\rightarrow C_-(\lam)/G_{p}C_-(\lam)\rightarrow
 C_-(\lam)/G_{p+1}C_-(\lam)\rightarrow 0,
\end{align*}
where $p\geq -1$.
This induces the exact sequence
\begin{align*}
 0\rightarrow G^{p+1}C_-(\lam)^*
\rightarrow
G^p C_-(\lam)^*
\rightarrow \left( G_{p+1}C_-(\lam)/G_{p}C_-(\lam)\right)^*
\rightarrow 0
\end{align*}
Therefore
\begin{align}\label{eq:10-21-dual-fil}
 \begin{aligned}
   G^p C_-(\lam)^*/G^{p+1}C_-(\lam)^*&=
\left( G_{p+1}C_-(\lam)/G_{p}C_-(\lam)\right)^*\\
&=
D\left( G_{p+1}C_-(\lam)/G_{p}C_-(\lam)
\right).
 \end{aligned}
\end{align}
Here,
 the last equality follows from  Lemma \ref{Lem:GpC-is-submodule}
(2).
Again by Lemma \ref{Lem:GpC-is-submodule} (2),
we have the following proposition.
\begin{Pro}
\label{Pro:duality-gr}
We have
 \begin{align*}
  H^{i}(G^p C_-(\lam)^*/G^{p+1}C_-(\lam)^*
)
=D(H^{-i}(G_{p+1} C_-(\lam)/G_p C_-(\lam)
)
 \end{align*}
for each $i$ and $p$.
\end{Pro}
\begin{Rem}\label{Rem:remark-of-red-coh}
 It is not the case that $H^i(C_-(\lam)^*)=D(H^{-i}(C_-(\lam)))$.
\end{Rem}
\subsection{}
Consider the 
subcomplex
\begin{align*}
 \gr^G C_-(\lam)_{\xi_{\lam}}
\subset \gr^G C_-(\lam)\teigi \sum_p G_p C_-(\lam)/G_{p-1}C_-(\lam).
\end{align*}By definition,
we have
\begin{align*}
 \gr^G C_-(\lam)_{\xi_{\lam}}
=\bigoplus_p
G_p C_-(\lam)_{\xi_{\lam}}/G_{p-1}C_-(\lam)_{\xi_{\lam}},
\end{align*}and
$d_-^{\chi}$
acts trivially on  $ \gr^G C_-(\lam)_{\xi_{\lam}}$
(see \eqref{eq:def-of-Gp-bar}).
Thus,
\begin{align}
  (\gr^G
C_-(\lam)_{\xi_{\lam}},d_-)=(C_-(\lam)_{\xi_{\lam}},d_-^{\st})
\end{align}
as complexes.
In particular,
\begin{align}\label{eq:ss-sq-of-hw}
(E_1^{\bullet,q})_{\xi_{\lam}}
=H^{q}(\gr^G C_-(\lam)_{\xi_{\lam}}^*,d)
=H^{q}(C_-(\lam)_{\xi_{\lam}}^*,d^{\st}),
\end{align}
because
the complex
$(C_-(\lam)_{\xi_{\lam}},d_-^{\st})$
is a direct sum of finite-dimensional
subcomplexes $(C_-(\lam)_{\xi_{\lam}}^{\mu},d_-^{\st})$,
with $\mu \in \dual{\bh}$.
\subsection{}
Consider the complex
\begin{align*}
 G_0 C_-(\lam)/G_{-1}C_-(\lam)=C_-(\lam)/G_{-1}C_-(\lam).
\end{align*}
Let
$\overline{|\lam\ket}$
be the image of $|\lam\ket_-$
in $C_-(\lam)/G_{-1}C_-(\lam)$.
Then,
the following hold:
\begin{align}
& (C_-(\lam)/G_{-1}C_-(\lam))_{\xi_{\lam}}=\C \overline{|\lam\ket},\nonumber\\
 &J^{(v)}_-(n)\overline{|\lam\ket}=0\quad 
\text{for }
v(n)\in L\sg_{\geq 0}\cap \bg_+,\nonumber\\
&\psi^{-\alpha}(n)\overline{|\lam\ket} =0\quad 
\text{for }\alpha\in \sroots_{>0},
n\geq 0,\nonumber\\
&\Phi_{-\alpha}(n)\overline{|\lam\ket}=0\quad
\text{for } \alpha\in \sroots_{\frac{1}{2}},
n\geq 1,\nonumber\\
& J^{(e)}(-1)\overline{|\lam\ket}
=\psi^{-\theta}(-1)\overline{|\lam\ket}=0,
\\
&J^{(h)}_-(0)\overline{|\lam\ket} =\bra \lam,h\ket \overline{|\lam\ket}\quad
\text{for }h\in \sh.\nonumber
\end{align}

\begin{Lem}
\label{Lem:dec-of-comp1}
  We have
\begin{align*}
\gr^G C_-(\lam)
=\bigoplus\limits_{\mu\in \dual{\bh}}
\left(C_-(\mu)/G_{-1}C_-(\mu)\right)
\*  C_-(\lam)_{\xi_{\lam}}^{\mu}
\end{align*}as complexes,
where 
$C_-(\lam)_{\xi_{\lam}}^{\mu}=(C_-(\lam)_{\xi_{\lam}}^{\mu},d_-^{\st})$.
\end{Lem}
\begin{proof}
Define a linear map 
$\bigoplus\limits_{\mu\in \dual{\bh}}
 \left(C_-(\mu)/G_{-1}C_-(\mu)\right) 
\* \gr^G C_-(\lam)^{\mu}_{\xi_{\lam}}
\rightarrow   \gr^G C_-(\lam)
$ by
\begin{align*}
& J^{(u_1)}_-(m_1)\dots J^{(u_p)}_-(m_p)
\Phi_{-\alpha_1}(n_1)\dots \Phi_{-\alpha_q}(n_q)
\psi^{-\beta_1}(s_1)\dots \dots \psi^{-\beta_r}(s_r)
\overline{|\mu\ket}\* v
\\
&\mapsto
 J^{(u_1)}_-(m_1)\dots J^{(u_p)}_-(m_p)
\Phi_{-\alpha_1}(n_1)\dots \Phi_{-\alpha_q}(n_q)
\psi^{-\beta_1}(s_1)\dots \dots \psi^{-\beta_r}(s_r)v,
\end{align*}
where $u_i\in \sg_{\geq 0}$,
$\alpha\in \sroots_{\frac{1}{2}}$,
$\beta_i\in \sroots_{>0}$,
$m_i,n_i,s_i\in \Z$ and
$v\in \gr^G C_-(\lam)_{\xi_{\lam}}^{\mu}$.
Then, it can be verified
this is an isomorphism of complexes
({cf.\ Proposition 6.12 of Ref.\ \cite{A1}}).
\end{proof}
\subsection{}
Let
\begin{align*}
& \left(\bge\right)_-\teigi 
\sn_{0,-}\* \C[t\inv]\+
(\sh^e\+ \sn_{0,+}\+\sg_{\frac{1}{2}})\* \C[t\inv]\inv
\+ \sg_1\*\C[t\inv]t^{-2}.
\end{align*}
The proof of the following assertion is the same as
that of Theorem 4.1 of Ref.\ \cite{KW2003}.
\begin{Pro}
 \label{Pro:van-bar-10}
For any $\lam\in \dual{\bh}$,
we have
 $H^i(C_-(\lam)/G_{-1}C_-(\lam))
=\{0\}$ for  $i\ne 0$
and the following map defines an isomorphism
of $\C$-vector spaces:
\begin{align*}
 \begin{array}{ccc}
 U((\bge)_-) &\rightarrow  & H^0(C_-(\lam)/G_{-1}C_-(\lam)), \\
u_1(n_1)\dots u_r(n_r)&\mapsto & W_-^{(u_1)}(n_1)\dots
W_-^{(u_r)}(n_r)\overline{|\lam\ket}.
 \end{array}
\end{align*}
\end{Pro}

Now recall that $ G^{-1}C_-(\lam)^*
=D(C_-(\lam)/G_{-1}C_-(\lam))\subset C_-(\lam)^*
$
(see \eqref{eq:10-21-dual-fil}).
Because $G^{-1}C_-(\lam)^*$
is a $C_k(\sg)$-submodule of $C_-(\lam)^*$,
it follows that 
$H^{\bullet}( G^{-1}C_-(\lam)^*)$ 
is a module over   $\W_k(\sg,f)$.
It is clear that
\begin{align}\label{eq:11-2-wt-sp}
 &H^{\bullet}( G^{-1}C_-(\lam)^*)=\bigoplus_{\xi\leq \xi_{\lam}}
H^{\bullet}( G^{-1}C_-(\lam)^*)_{\xi}
\text{ and }H^{\bullet}( G^{-1}C_-(\lam)^*)_{\xi_{\lam}}=
\C |\lam\ket^*.
\end{align}
Here with a slight  abuse of notation
we have denoted  
 the  vector in $H^{\bullet}( G^{-1}C_-(\lam)^*)_{\xi_{\lam}}$
 dual to $\overline{|\lam\ket}$ by $|\lam\ket ^*$.
\begin{Pro}
\label{Pro:van-and-cofree-10}
$ $

\begin{enumerate}
 \item $H^i(G^{-1}C_-(\lam)^*)=\{0\}$ for $i\ne 0$.
 \item Any nonzero $\W_k(\sg,f)$-submodule of 
 $H^0(G^{-1}{C_-(\lam)}^*)$
contains the canonical vector 
${|\lam\ket}^*\in H^0(G^{-1}C_-(\lam)^*)_{\xi_{\lam}}$.
\end{enumerate} 
\end{Pro}
\begin{proof}
 (1) 
It is 
 straightforward to demonstrate using
 Propositions \ref{Pro:duality-gr}
and   \ref{Pro:van-bar-10}.
(2)  
For a $\W_k(\sg,f)$-module $M$,
let $\HW(M)$
be the space of  singular vectors:
\begin{align*}
\HW(M)\teigi   \{ m\in M\mid\
W^{(u)}(n)m=0\text{ for all $v(n)\in (\bgf)_+$}\}\subset M.
\end{align*}
Then, by \eqref{eq:11-2-wt-sp},
we have $\HW(M)\ne \{0\}$
for  any nonzero $\W_k(\sg,f)$-submodule $M$ of  $H^0(G^{-1}C_-(\lam)^*)$.
Hence,
 it is sufficient to show that 
\begin{align}\label{Hq-eq;vector}
 \HW(H^0( G^{-1}C_-(\lam)^*))= \C |\lam\ket^*.
\end{align}
Note that,
by \eqref{eq:11-2-wt-sp},
it is obvious that
$ \HW(H^0(G^{-1}C_-(\lam)^*))\supset 
\C  |\lam\ket^*$.
Therefore,
 we have only to 
show 
\begin{align}\label{eq:11-04-inclusion}
 \HW(H^0(G^{-1}C_-(\lam)^*))\subset 
\C  |\lam\ket^*.
\end{align}

To demonstrate \eqref{eq:11-04-inclusion},
we make use of
 the filtration $\{F_p\W_k(\sg,f)\}$
of $\W_k(\sg,f)$
given in Theorem \ref{th:str-of-W},
described by \eqref{eq:description-of-filtration}.
First, set $ F_{-1} H^{0}(C_-(\lam)/G_{-1}C_-(\lam))=\{0\}$
and
\begin{align*}
& F_p H^{0}(C_-(\lam)/G_{-1}C_-(\lam))\\
&\teigi 
\haru\{ W^{(u_1)}_-(n_1)\dots W^{(u_r)}_-(n_r)\overline{|\lam\ket}\mid 
u_i\in \sge_{s_i},\
\sum s_i\leq  p\} \quad \text{for }p\geq 0.
\end{align*}
Next, define
\begin{align*}
& F_pH^0(G^{-1}C_-(\lam)^*)\\
&=D\left(
H^0(C_-(\lam)/G_{-1}C_-(\lam))/
F_{-p}H^0(C_-(\lam)/G_{-1}C_-(\lam))\right)
\quad \text{for }p\leq 1.
\end{align*}
(Recall that
$H^0(G^{-1}C_-(\lam)^*)=D(H^0(C_-(\lam)/G_{-1}C_-(\lam)))$).
Then,
we have
\begin{align}
  &
\begin{aligned}
 &\dots \subset F_p H^0(G^{-1}C_-(\lam)^*)
\subset 
F_{p+1} H^0(G^{-1}C_-(\lam)^*)\subset \dots  \\
&\quad \dots
\subset   F_0H^0(G^{-1}C_-(\lam)^*) \subset 
F_1H^0(G^{-1}C_-(\lam)^*)
=H^0(G^{-1}C_-(\lam)^*),\\
\end{aligned}\\
& \bigcap F_p H^0(G^{-1}C_-(\lam)^*)=\{0\},\\
& F_p\W_k(\sg,f)\cdot F_q H^0(G^{-1}C_-(\lam)^*)
\subset F_{p+q} H^0(G^{-1}C_-(\lam)^*), \label{eq:10-comp-fil}
\end{align}
where $ F_p\W_k(\sg,f)\cdot F_q H^0(G^{-1}C_-(\lam)^*)
$
denotes
the span of the vectors 
$Y_n(v)w$ with $v\in F_p \W_k(\sg,f),\ w\in F_q H^0(G^{-1}C_-(\lam)^*)$
 and
$ n\in
 \Z$.

By \eqref{eq:10-comp-fil},
the corresponding
 graded
space $\gr_F H^0(G^{-1}C_-(\lam)^*)$,
is a module 
over
$\gr_F \W_k(\sg,f)=V_k^{\natural}(\sgf)$.
In particular,
$\gr_F H^0(G^{-1}C_-(\lam)^*)$ is a module over 
the Lie algebra 
$(\bgf)_+$.
(Observe $(\bgf)_+$ is generated by the image of $W^{(u)}(n)$,
$u(n)\in (\bgf)_+$.)
Moreover,
 from (the proof of)
Proposition \ref{Pro:van-bar-10},
it follows that
\begin{align}\label{eq:iso-fr-10}
 \gr_F H^0(G^{-1}C_-(\lam)^*)\cong D(U(\bge)_-)
\quad \text{as $(\bgf)_+$-modules,}
\end{align}
where $(\bgf)_+$ acts on $D(U(\bge)_-)$
by $(u(n)g)(v)=g(u^t(-n)v)$,
where  $u\in \sgf$, $v\in U(\bge)$,
$g\in D(U(\bge))$.
Hence,
it follows that
\begin{align*}
   H^0((\bgf)_+, \gr_F H^0(G^{-1}C_-(\lam)^*))
= \C (\text{the image of $|\lam\ket^*$}).
\end{align*}
But this implies that
$
 \HW(H^0(G^{-1}C_-(\lam)^*)))\subset 
\C  |\lam\ket^*
$.
This completes the proof.
\end{proof}
\subsection{}
\begin{Th}\label{Th:verma-in-cofree-proof}
  Suppose that
$\bra \lam,\alpha_0\che\ket\not\in \{0,1,2,\dots\}$.
Then,
$H^{i}(M(\lam)^*)=\{0\}$
for
$i\ne 0$
and $H^0(M(\lam)^*)=H^0(G^{-1}C_-(\lam)^*)$
as $\W_k(\sg,f)$-modules.
In particular,
any nonzero $\W_k(\sg,f)$-submodule of 
 $H^0(M(\lam)^*)$
contains the canonical vector ${|\lam\ket}^*$.
\end{Th}
\begin{proof}
We have 
$\bar{M}_{\sl_2}(\bra \lam,\alpha_0\che\ket)^*=
\bar{M}_{\sl_2}(\bra \lam,\alpha_0\che\ket)$
by assumption.
Also,
we have $H^i(C_-(\lam)_{\xi_{\lam}},d^{\st}_-)
\cong H^{i}
(\C e,\bar{M}_{\sl_2}(\bra \lam,\alpha_0\che\ket))$.
This can be demonstrated in the same manner as Lemma
 \ref{Lem:cohomology-of-zero-mode}.
Hence, it follows that
\begin{align*}
 H^i(C_-(\lam)_{\xi_{\lam}}^{\mu},d^{\st}_-)
=\begin{cases}
 \C&(i=0\text{ and } \mu=\lam), \\
\{0\}&\text{(otherwise)}.
 \end{cases}
\end{align*}
Therefore, by Lemma \ref{Lem:dec-of-comp1} and 
Proposition \ref{Pro:van-bar-10},
we have
\begin{align*}
 H^i(\gr^G C_-(\lam))=\begin{cases}
		       H^0(C_-(\lam)/G_{-1}C_-(\lam))&(i=0),\\
\{0\}&(i\ne 0).
		      \end{cases}
\end{align*}
Hence, by Proposition \ref{Pro:duality-gr},
we have
\begin{align*}
  H^i(\gr^G C_-(\lam)^*)=\begin{cases}
			 H^0(G^{-1}C_-(\lam)^*)&(i=0)\\
\{0\}&(i\ne 0).
			\end{cases}
\end{align*}
This implies that
our spectral sequence collapses at $E_1=E_{\infty}$,
and therefore
\begin{align*}
 H^i(M(\lam)^*)
=\begin{cases}
			 H^0(G^{-1}C_-(\lam)^*)
&(i=0)\\
\{0\}&(i\ne 0)
			\end{cases}
\end{align*}
as vector spaces.
But the isomorphism
$ H^0(G^{-1}C_-(\lam)^*)\cong H^0(M(\lam)^*)$
is induced by the $C_k(\sg)$-module
homomorphism $G^{-1}C_-(\lam)^*\hookrightarrow C_-(\lam)^*$,
and
hence it is a $\W_k(\sg,f)$-homomorphism.
\end{proof}
\subsection{}\label{subsection:proof-of-vanishing-of-dual-of-Verma}
We finally consider the case in which $\lam$ is a general element
of $\dual{\bh}_k$.
By Proposition \ref{Pro:van-bar-10},
we have
\begin{align}\label{eq:identification-art}
 H^0( G^{-1}C_-(\mu)^*)
\cong H^0(G^{-1}C_-(\mu')^*)
\end{align}
as $\C$-vector spaces for any $\mu,\mu'\in \dual{\bh}$.
With 
\eqref{eq:identification-art},
it follows from  \eqref{eq:E-1-first-identity},
Proposition  \ref{Pro:duality-gr}
and  Lemma \ref{Lem:dec-of-comp1}
that
we have the isomorphism 
\begin{align}\label{eq:identity-iso-cocyclic}
\begin{aligned}
   E_1^{\bullet,q}&\cong 
H^0(G^{-1}C_-(\lam)^*)
\*  H^q(
C_-(\lam)^*_{\xi_{\lam}},d^{\st})\\
&=H^0(G^{-1}C_-(\lam)^*)\*
(E_1^{\bullet,q})_{\xi_{\lam}}
\end{aligned}\end{align}
of complexes,
where
the differential $d_1$ acts on the first factor
$H^0(G^{-1}C_-(\lam)^*)$ trivially.
%
This 
induces the isomorphisms
\begin{align}
 (E_r,d_r)\cong (H^0(G^{-1}C_-(\lam)^*)\* (E_r)_{\xi_{\lam}},
1\* d_r)
\end{align}
inductively  for all $r\geq 1$.
Therefore,
we obtain
\begin{align}\label{eq:final-id-of-sc}
 E_{\infty}\cong H^0(G^{-1}C_-(\lam)^*)
\* (E_{\infty})_{\xi_{\lam}}.
\end{align}
On the other hand,
by \eqref{eq:convergenceof-sp-seqence-04},
we have
\begin{align*}
  (E_{\infty})_{\xi_{\lam}}=H^{\bullet}(M(\lam)^*)_{\xi_{\lam}}.
\end{align*}
Hence, by 
\eqref{eq:final-id-of-sc}
and  Proposition \ref{Pro:theVanishin-zeromode-1} (3),
we have 
proved
Theorem \ref{Th;vanishing-and-cocyclic-dual-2},
as desired.
\renewcommand{\qedsymbol}{$\blacksquare$}
\qed
\renewcommand{\qedsymbol}{$\square$}


\appendix

\section{The setting for $A(1,1)$}
In this appendix we summarize the change of the setting
 for the $A(1,1)$ case.
\subsection{The  setting for $\sg$}
Let $\sg=\sl(2|2)/\C I$
and
let $\sh'$ be the Cartan subalgebra 
of $\sg$ containing $x$.
Let 
 $\sa\teigi \gl(2|2)/\C I$.
Then,
$\sg=[\sa,\sa]\subset \sa$.
Let $\sh$ be
 the Cartan subalgebra of $\sa$
containing $\sh'$.
(So $\dim \sh=3$.)
 Then,
$[\sh,\sg]=\sg$
and we have
 $\sg=\bigoplus_{\alpha\in \dual{\sh}}\sg^{\alpha}$,
where $\sg^{\alpha}=\{ u\in \sg\mid
[h,u]=\bra \alpha,h\ket u\text{ for all $h\in \sh$}\}$.
Define   the set $\sroots$
as the subset of $\dual{\sh}$
consisting of elements $\alpha$
such that $\sg^{\alpha}\ne \{0\}$.
Then,
$\dim \sg^{\alpha}=1$ for all $\alpha\in \sroots$.
The remaining  setting is  the same as in the other cases.
\subsection{The  setting for $\bg$}
Let $\bg$ be the  affine Lie algebra 
associated with  $\sg=\sl(2|2)/\C I$
defined 
by \eqref{eq-def-of-affine-Lie-algebra}.
Let
$\bg=\bn_-\+ \bh'\+ \bn_+$
be the standard triangular decomposition,
where $\bh'=\sh' \+ \C K \+ \C \Dg$ is the standard Cartan subalgebra of $\bg$.
Next, define
the commutative Lie algebra $\bh$
by
 $\bh\teigi \sh\+ \C K\+ \C \Dg$,
where $\sh$ is  as above.
Then, the action of $\sh$ on $\sg$
naturally extends to the action of $\bh$ 
on $\bg$.
This 
gives
the space
\begin{align}\label{eq:widetilde{sg}}
 \widetilde{\sg}\teigi \bn_-\+ \bh\+ \bn_+
\end{align}
a Lie superalgebra structure such that
$\bg= [\widetilde{\sg},\widetilde{\sg}]$,
and $\bh$ is a Cartan subalgebra of $\widetilde{\sg}$.
Now define $\broots\subset \dual{\bh}$
as the set of roots of $\widetilde{\sg}$
and
$\broots_+\subset \broots$
as the set of positive roots of $\widetilde{\sg}$
(according to the triangular decomposition \eqref{eq:widetilde{sg}}).
Also,
we define $\bQ=\sum_{\alpha\in \roots}\Z \alpha\subset
\dual{\bh}$, $\bQ_+=\sum_{\alpha\in \broots_+}\Z_{\alpha}\subset
\bQ$
and a partial ordering $\mu\leq \lam$ 
on $\dual{\bh}$
by $\lam-\mu\in \bQ_+$.

Next, 
we replace $\bg$-modules by $\widetilde{\sg}$-modules.
In particular,
we define  the category $\BGG_k$
as the full  subcategory of the category
of $\widetilde{\sg}$-modules satisfying the conditions of 
Subsection \ref{subsection:category-O}.
Note
that
the simple $\widetilde{\sg}$-module $L(\lam)$,
with $\lam\in \dual{\bh}$,
is already irreducible as a $\bg$-module.
This fact can be seen using the 
argument of the proof of Theorem \ref{Th;image-of-simples}.
The remaining  setting is the same as in the other cases.


\begin{thebibliography}{9}
\bibitem{A1}
{Arakawa, T.}:
{\em Vanishing of Cohomology Associated to Quantized
Drinfeld-Sokolov Reduction},
Int.\ Math.\ Res.\ Notices, 2004, No.\ 15,
729--767.
\bibitem{A2}
{Arakawa, T.}:
{\em Quantized Reductions and Irreducible Representations of 
$\W$-Algebras},
preprint, math.QA/0403477.
\bibitem{BGG}
{Bernstein, I. N.,  Gel'fand, I. M.,  Gel'fand, S. I. }:
{\em A certain category of $\g$-modules.}
Functional.\ Anal.\ Appl.\ 10 (1976), 87--92.
\bibitem{BeK}{Bershadsky, M.}:
Phys.\ Lett.\ 174B (1986) 285;
{Knizhnik, V.G.}:
Theor.\ Math.\ Phys.,
66 (1986) 68.
\bibitem{B}{Bershadsky, M.}:
{\em Conformal field theory via Hamiltonian reduction},
Comm.\ Math.\ Phys.\ 139 (1991) 71-82.
\bibitem{D}
{Dobrev, V. K.}:
{\em Characters of the unitarizable
 highest weight modules over the $N=2$ superconformal algebras},
Phys.\ Lett.\ B 186 (1987), no.\ 1, 43--51.
\bibitem{FL}
{Fateev, V. A.,  Lukyanov, S. L.}:
{\em The models of two-dimensional conformal quantum field theory with 
$Z_n$ symmetry. }
Internat.\ J. Modern Phys. A 3 (1988), no. 2, 507--520.
\bibitem{FZa}
{Fateev, V. A., Zamolodchikov, A. B.};
{\em Conformal quantum field theory models in two dimensions having 
$Z_3$ symmetry. }
Nuclear Phys. B 280 (1987), no. 4, 644--660.
\bibitem{Feigin}
{Fe\u{\i}gin, B. L.}:
Semi-infinite homology of Lie, Kac-Moody and Virasoro algebras.
Uspekhi Mat.~Nauk 39 (1984), no. 2(236), 195--196.
\bibitem{FF_W} {Fe\u{\i}gin, B. L., Frenkel, E. V.}:
{\em Quantization of Drinfeld-Sokolov Reduction},
Phys.\ Lett.\ B 246, 75-81 (1990).
\bibitem{FF_W2} {Fe\u{\i}gin, B. L., Frenkel, E. V.}:
{\em Affine Kac-Moody algebras at the critical level
and Gel'fand-Diki\v{\i} algebras},
Int.\ Jour.\ Mod.\ Phys.\
A7, Supplement 1A (1992) 197-215.
\bibitem{FB} {Frenkel, E. V.,  Ben-Zvi, D.}: {\em Vertex algebras and
algebraic curves},
 Mathematical Surveys and Monographs, 88.
\bibitem{FKW} {Frenkel, E. V., Kac, V. G., Wakimoto, M.}:
{\em  Characters
and fusion rules for
$W$-algebras via quantized Drinfeld-Sokolov reduction},
Comm.\ Math.\ Phys.\ 147 (1992), no. 2, 295--328.
\bibitem{FZ}
{Frenkel, I.\  B., Zhu, Y.}:
{\em Vertex operator algebras associated to representations 
of affine and Virasoro algebras}. 
Duke Math.\ J. 66 (1992), no.\ 1, 123--168.
\bibitem{KacBook}
{Kac, Victor G.}:
{\em Infinite-dimensional {L}ie algebras},
{Cambridge University Press},
{1990}.
\bibitem{KacLNM}
{Kac, Victor G.}:
{\em Representations of classical Lie superalgebras}.
Differential geometrical methods in mathematical physics, II 
(Proc. Conf., Univ. Bonn, Bonn, 1977), 597--626, 
Lecture Notes in Math., 676, 
Springer, Berlin, 1978.
\bibitem{KRW}
{Kac, V.\ G., Roan, S.-S., Wakimoto, M.}: 
{\em Quantum Reduction for Affine Superalgebras},
Commun.\ Math.\ Phys.\ 241 (2003),
307-342.
\bibitem{KW1} 
{Kac, V.\ G., Wakimoto, M.}: 
{\em Modular invariant
representations of infinite-dimensional Lie algebras and
superalgebras}. Proc.\ Nat.\ Acad.\ Sci.\ U.S.A.\ 85 (1988), no.\ 14,
4956--4960.
\bibitem{KW2003}
{Kac, V.\ G., Wakimoto, M.}: 
{\em Quantum Reduction amd Representation Theory of
Superconformal Algebras},
Adv.\ in Math.\ 185 (2004) 400-458.
\bibitem{KW2004-2}
{Kac, V.\ G., Wakimoto, M.}: 
{\em Quantum reduction and characters of superconformal algebras},
to appear.
\bibitem{KT}
{Kashiwara, M., Tanisaki, T.};
{\em Characters of irreducible modules
with non-critical highest weights over affine Lie algebras}.
Representations and quantizations (Shanghai, 1998),
275--296,  China High.\ Educ.\ Press, Beijing, 2000.
\bibitem{Kh}
{Khovanova, T,}:
{\em Super KdV equation related to the Neveu-Schwarz-2 Lie superalgebra
of string theory. }
Theor.\ Mat.\ Phys.\ 72, 306-312 (1987).
\bibitem{Ki}
{Kiritsis, Elias B}:
{\em Character formulae 
and the structure of the representations 
of the $N=1,\;N=2$ superconformal algebras}. 
Internat.\  J.\  Modern Phys.\  A 3 (1988), no\  8, 1871--1906.



\bibitem{Luk}
{Lukyanov, S. L.}:
{\em Quantization of the Gel'fand-Dikii bracket, } 
Funktsional. Anal. i Prilozhen. 22 (1988), no. 4, 1--10, 96; 
translation in Funct. Anal. Appl. 22 (1988), no. 4, 255--262 (1989)
\bibitem{LF}
{Lukyanov, S. L., Fateev, V. A.};
{\em Conformally invariant models of two-dimensional quantum field
	theory with 
$Z_n$-symmetry, }
Soviet Phys. JETP 67 (1988), no. 3, 447--454.
\bibitem{M}
{Matsuo, Y.}:
{\em Character formula of $c<1$ unitary representation 
of $N=2$ superconformal algebra}. 
Progr.\  Theoret.\  Phys.\  77 (1987), no.\  4, 793--797.

\bibitem{MPBOOK}
{Moody, R.\ V., Pianzola, A.}
{\em Lie algebras with triangular decompositions},
Canadian Mathematical Society Series of Monographs and Advanced Texts.
A Wiley-Interscience Publication.
John Wiley \& Sons, Inc., New York, 1995.
\bibitem{RY}
{Ravanini, F. and  Yang, S.-K.}:
{\em Modular invariance in $N=2$ superconformal field theories}. 
Phys.\ Lett.\ B 195 (1987), no.\ 2, 202--208.



\bibitem{Za}{Zamolodchikov, A}:
{\em Infinite additional symmetries in two-dimensional conformal
field theory},
Theor.\ Math.\ Phys.\ 65 (1985), 1205--1213
\end{thebibliography}
\end{document}